\documentclass[a4paper,11pt]{article}
\pdfoutput=1 

\usepackage{jcappub} 

\usepackage[T1]{fontenc} 

\def\lsim{\mathrel{\rlap{\lower4pt\hbox{\hskip1pt$\sim$}} \raise1pt\hbox{$<$}}}
\def\gsim{\mathrel{\rlap{\lower4pt\hbox{\hskip1pt$\sim$}} \raise1pt\hbox{$>$}}}
\usepackage{hyperref}
\usepackage{graphicx, color}
\usepackage{xcolor}
\usepackage[export]{adjustbox}
\usepackage{cancel}

\usepackage[normalem]{ulem}

\newcommand{\newc}{\newcommand}
\newc{\comment}[1]{}

\definecolor{grassgreen}{cmyk}{0.77,0,1,0.05}

\newcommand{\be}{\begin{equation}}
\newcommand{\ee}{\end{equation}}
\newcommand{\bea}{\begin{equation}\begin{aligned}}
\newcommand{\eea}{\end{aligned}\end{equation}}

\title{Composite Hybrid Inflation : Primordial Black Holes and Stochastic Gravitational Waves}

\author[a,b]{Giacomo Cacciapaglia,}
\affiliation[a]{Laboratoire de Physique Th\'eorique et Hautes \'Energies (LPTHE), UMR 7589,\\
Sorbonne Universit\'e \& CNRS, 4 place Jussieu, 75252 Paris Cedex 05, France}
\affiliation[b]{Quantum Theory Center ($\hslash$QTC) at IMADA \& D-IAS, Southern Denmark Univ., Campusvej 55, 5230 Odense M, Denmark}
\emailAdd{cacciapa@lpthe.jussieu.fr}

\author[c]{Dhong Yeon Cheong,}
\affiliation[c]{Department of Physics and IPAP, Yonsei University, Seoul 03722, Republic of Korea}
\emailAdd{dhongyeon@yonsei.ac.kr}

\author[d,e]{Aldo Deandrea,}
\affiliation[d]{Universit\'e Claude Bernard Lyon 1, CNRS/IN2P3, IP2I UMR 5822, 4 rue Enrico Fermi, F-69100 Villeurbanne, France}
\affiliation[e]{Department of Physics, University of Johannesburg, PO Box 524, Auckland Park 2006, Johannesburg, South Africa}
\emailAdd{deandrea@ip2i.in2p3.fr}

\author[d]{Wanda Isnard,}
\emailAdd{w.isnard@ip2i.in2p3.fr}

\author[c,f]{Seong Chan Park,}
\affiliation[f]{Korea Institute for Advanced Study, Seoul 02455, Republic of Korea}
\emailAdd{sc.park@yonsei.ac.kr}

\author[g,h,i]{Xinpeng Wang,}
\emailAdd{xinpeng.wang@ipmu.jp}
\affiliation[g]{Kavli Institute for the Physics and Mathematics of the Universe (WPI), The University of Tokyo Institutes for Advanced Study, The University of Tokyo, Chiba 277-8583, Japan}
\affiliation[h]{School of Physics Science and Engineering, Tongji University, Shanghai 200092, China}
\affiliation[i]{Department of Physics, Graduate School of Science, The University of Tokyo, Tokyo 113-0033, Japan}

\author[h,j,g,k,l]{Ying-li Zhang}
\emailAdd{yingli@tongji.edu.cn}
\affiliation[j]{Institute for Advanced Study of Tongji University, Shanghai 200092, China}
\affiliation[k]{Institute of Theoretical Physics, Chinese Academy of Sciences, Beijing 100190, China}
\affiliation[l]{Center for Gravitation and Cosmology, Yangzhou University, Yangzhou 225009, China}

\abstract{We investigate the production of primordial black holes and gravitational waves in composite hybrid inflation. Starting from an effective chiral Lagrangian with a dilaton and pions, we identify inflation occurring due to the walking dynamics of the theory. A $\mathbb{Z}_2$ symmetry-breaking term in the pion sector induces a shift in the inflaton's trajectory, which leads to a tachyonic instability phase. Curvature perturbations grow exponentially, producing copious primordial black holes and a stochastic gravitational wave background. We show that the primordial black hole mass and the gravitational wave frequency are strongly restricted by the anomalous dimensions of the pion operators, with larger anomalous dimensions giving lighter primordial black holes and higher frequency gravitational waves. In both cases, the associated signatures lie within reach of future gravitational wave observatories. }

\begin{document}
\maketitle
\flushbottom

\section{Introduction}
\label{sec:intro}

Cosmological inflation is currently the best motivated and developed framework to describe the initial conditions of our early universe. It naturally solves the flatness and horizon problems, and provides a mechanism to generate the primordial perturbations leading to the structure formation of our universe \cite{Starobinsky:1980te,Guth:1980zm,Linde:1981mu,Starobinsky:1982ee,Linde:1983gd,Linde:2007fr,Baumann:2008bn,Kinney:2009vz,Martin:2018ycu}. Inflation could also provide a way to access energy densities far beyond the reach of particle accelerators. However, the microscopic origin of the inflation mechanism remains to be established, and the search for a complete feasible model remains an active field of research (see e.g.~\cite{Martin:2013tda} for a comprehensive review of inflationary models). Among many possibilities, models of composite inflation have already been considered in the past, see \cite{Channuie_2015,2022EPJST.231.1325S,Liu:2024xrh}. For a review of phenomenological applications of composite dynamics, including inflation, we refer the reader to \cite{Cacciapaglia:2022zwt}.

In this paper, we further investigate the imprints of a composite hybrid inflation model started in \cite{Cacciapaglia:2023kat}. We consider a general framework of confining gauge theory coupled to fundamental fermions. In the confined phase, the spontaneous breaking of the chiral and scale symmetries leads to the presence of light composite scalars: pions and a dilaton, respectively. The dilatonic potential is constrained by the scale anomaly and it naturally contains a flat direction that can be used to drive inflation in the early universe \cite{Golterman:2016lsd,Ishida:2019wkd,Appelquist:2017wcg}. Instead, the pion potential is due to the interactions of the fundamental fermions, and it provides a way to end inflation via a waterfall direction \cite{Cacciapaglia:2023kat}. The most minimal pion potential, however, features two non-equivalent vacua, due to an intrinsic $\mathbb{Z}_2$ symmetry. Compared to the previous analysis, in this work the chiral Lagrangian leading to the pion potential is further expanded to include a $\mathbb{Z}_2$ breaking term, in such a way that the degeneracy between the different vacua is lifted and domain walls decay before they can dominate the energy density of the universe. The decay of those domain walls leads to high frequency gravitational waves (GW) signatures, which could {potentially be accessible} by ultra-high-frequency GW observatories~\cite{Ejlli:2019bqj,Ito:2019wcb,Ito:2020wxi,Aggarwal:2020olq, Ito:2022rxn,Ito:2023nkq,Berlin:2023grv,Aggarwal:2025noe}, see Appendix~\ref{app:domainwalls} for more details. 

Furthermore, the $\mathbb{Z}_2$ breaking term also induces a shift in the inflaton's trajectory leading to a tachyonic instability phase, which is the main focus of our work. We show that it yields an enhancement of the curvature perturbations leading to the production of primordial black holes (PBH) \cite{10.1093/mnras/152.1.75,10.1093/mnras/168.2.399,10.1093/mnras/215.4.575,Zeldovich:1967lct} and also of a stochastic GW background induced by scalar perturbations~\cite{Matarrese:1997ay,Mollerach:2003nq,Ananda:2006af,Baumann:2007zm,Saito:2008jc, Sasaki:2018dmp,Gong:2019mui,  Yuan:2021qgz,Domenech:2021ztg,Domenech:2019quo,Domenech:2020kqm,Cheong:2019vzl, Cheong:2022gfc,Braglia:2022phb,Wang:2024vfv,Iacconi:2024hmg,Kim:2025dyi,Wang:2025lti}. PBHs are a natural candidate for cold dark matter (CDM) as, for masses ranging from $10^{-16}~M_{\odot}$ to $10^{-12}~M_{\odot}$, they could constitute the whole observed dark matter density \cite{Niikura:2019kqi,Montero_Camacho_2019,Smyth_2020}. Two parameters of the confining dynamics play a crucial role, namely the anomalous dimensions associated to the operators that source the pion potential, $\gamma_x$. 
We find that, when the anomalous dimensions have values preferred by minimal compositeness models, $\gamma_x \le 1$, the PBHs produced at the end of the first stage of inflation are too massive to constitute the whole of dark matter. At the same time, the scalar-induced GW background (SIGW) can reach frequencies of nHz, which may correspond to the recent reports on an stochastic GW background through pulsar timing array (PTA) observatories~\cite{NANOGrav:2023gor,NANOGrav:2023hde,EPTA:2023fyk,Xu:2023wog,Miles:2024seg}.
We also consider larger anomalous dimensions, $1< \gamma_x \le 2$ as suggested in \cite{Ishida:2019wkd}: in this regime, the PBHs are in the mass range preferred by dark matter, with the associated SIGWs falling within reach of future GW observatories~\cite{LISA:2017pwj,Baker:2019nia, Seto:2001qf,Kawamura:2006up,Sato:2017dkf,Isoyama:2018rjb,Kawamura:2020pcg, Corbin:2005ny,Crowder:2005nr,Harry:2006fi, TianQin:2015yph, Ruan:2018tsw}.

The paper is organized as follows. We first detail the composite model and derive the potential for inflation in Section \ref{sec:model}. In Section \ref{sec:background}, we describe the background evolution and the different stages of inflation that arise from the potential. Then in Section \ref{sec:perturbation}, we compute the adiabatic and isocurvature perturbations that lead to a copious production of PBH and a stochastic GW background. In Section \ref{sec:gamma1}, we numerically compute and give quantitative results for the PBH and GW production by finding a suitable parameter space for inflation. Throughout the paper we choose the metric convention $(-,+,+,+)$, and fix the Planck mass $M_{P} = 1$, unless specified.

\section{The model}
\label{sec:model}

As a concrete example of fundamental dynamics for the composite theory, we consider a $SU(N_c)$ gauge theory coupling to $N_{f}$ {fundamental} Dirac fermions ~\cite{Cacciapaglia:2023kat}. The number of colors, $N_c$, and fermion multiplicity $N_f$ are chosen such that the theory { exhibits a phase of near scale invariant dynamics }~\cite{Yamawaki:1985zg, Holdom:1981rm} {due to the presence of} an infra-red fixed point \cite{Dietrich:2006cm} {in the renormalization flow}. {This is referred to as a walking dynamics, as the gauge coupling remains approximately constant over a wide range of energy scales.} This property allows for the emergence of scale invariance in the low energy limit of the theory. We further assume that some dynamics, such as for instance irrelevant couplings, seeds the formation of fermion and gluon condensates. They spontaneously break the scale and chiral symmetries, leading to the presence of pseudo Nambu-Goldstone bosons (pNGBs): the former breaking gives a dilaton $\chi$, while the latter leads to pions $\phi$. The theory, therefore, will be characterized by the scales $f_\chi$ and $f_\phi$, i.e. the decay constants of the dilaton and pions, respectively, which must be below the Planck scale for consistency. Hence, the inflationary physics must be described in terms of a low energy effective lagrangian containing only the lightest degrees of freedom of the theory. Note, finally, that this effective description applies to a large number of dynamical interactions.

\subsection{The inflationary potential}

The Lagrangian describing the dynamics of the dilaton and pions is obtained through the derivative expansion of the usual chiral Lagrangian \cite{Cacciapaglia:2014uja}, where the pion fields are represented through $U= \exp \left( i \phi / f_{\phi} \right)$ with $f_{\phi}$ the decay constant of the pions.  The dilaton, instead, couples to the various operators via their scaling dimension. The dilaton also obtains a potential through the scale anomaly. Including only the lowest order terms in the chiral expansion, we consider the following effective theory:
\bea
\frac{\mathcal{L}}{\sqrt{-g}} \supset & \frac{1}{2} R-\frac{1}{2} \partial_\mu \chi \partial^\mu \chi-\frac{f_\phi^2}{2}\left(\frac{\chi}{f_\chi}\right)^2 \operatorname{Tr}\left[\partial_\mu U^{\dagger} \partial^\mu U\right]-\frac{\lambda_\chi}{4} \chi^4\left(\log \frac{\chi}{f_\chi}-A\right) \\
& +\frac{\lambda_\chi \delta_1 f_\chi^4}{2}\left(\frac{\chi}{f_\chi}\right)^{3-\gamma_m} \operatorname{Tr}\left[U+U^{\dagger}\right]+\frac{\lambda_\chi \delta_2 f_\chi^4}{4}\left(\frac{\chi}{f_\chi}\right)^{2\left(3-\gamma_{4 f}\right)} \operatorname{Tr}\left[\left(U-U^{\dagger}\right)^2\right] \\
& + \frac{\lambda_\chi \delta_3 f_\chi^4}{2i}\left(\frac{\chi}{f_\chi}\right)^{3-\gamma_m} \operatorname{Tr}\left[U-U^{\dagger}\right] -V_0\,,
\eea
where the gravitational background is contained in the Ricci scalar $R$ and the determinant of the metric $g$.
The notations for the parameters of the theory are taken from \cite{Cacciapaglia:2023kat}: the breaking of scale symmetry is controlled by the coupling $\lambda_\chi$ in the dilaton potential (last term of the first line), which we take as an overall normalization factor for the potential. The terms in the second and third line stem from the pion potential, which is sourced by couplings in the underlying theory that break the chiral symmetry explicitly: for instance, mass terms and four-fermion interactions.  We extended the model in \cite{Cacciapaglia:2023kat} by an explicit $\mathbb{Z}_{2}$ breaking term, with its relative breaking size determined by the parameter $\delta_{3}$. Note that both $\delta_1$ and $\delta_3$ can be generated by fermion mass terms, hence they share the same mass anomalous dimension $\gamma_m$, while the term $\delta_2$ can be generated by four-fermion interactions, hence having a different anomalous dimension $\gamma_{4f}$.

Expanding each term, in the field space $(\chi, \phi)$,  we find a curved kinetic term metric $G$
\bea \label{eq:Gkin}
G(\chi, \phi) = \operatorname{diag}(1, \chi^2 / f_{\chi}^2 ),
\eea
and a scalar potential
\bea
V(\phi, \chi)= & -\lambda_\chi \delta_1 f_\chi^4\left(\frac{\chi}{f_\chi}\right)^{3-\gamma_m} \cos \frac{\phi}{f_\phi}-\lambda_\chi \delta_2 f_\chi^4\left(\frac{\chi}{f_\chi}\right)^{2\left(3-\gamma_{4 f}\right)} \sin ^2 \frac{\phi}{f_\phi} \\
&  -\lambda_\chi \delta_3 f_\chi^4\left(\frac{\chi}{f_\chi}\right)^{3-\gamma_m} \sin \frac{\phi}{f_\phi} +\frac{\lambda_\chi}{4} \chi^4\left(\log \frac{\chi}{f_\chi}-A\right)+V_0 .
\label{eq:potentialV}
\eea
Besides the two anomalous dimensions $\gamma_m$ and $\gamma_{4f}$, the potential contains $6$ free parameters, two scales and four coupling constants:
\begin{equation}
    (\delta_1,~\delta_2,~ \delta_3,~\lambda_{\chi},~f_{\chi},~f_{\phi}).
\end{equation}
In the following, we will fix the anomalous dimensions (which are technically determined by the confining dynamics), and vary the other parameters, including the field initial condition given by the $\chi$ value at the pivot scale $k^*=0.05$ Mpc$^{-1}$, denoted by $\chi^*$. Neglecting the small breaking $\delta_3$, the vacuum expectation value (VEV) of the pion field is $\phi_0~\sim~f_{\phi}~\arccos\left( \frac{\delta_1}{2 \delta_2} \right)$. By requiring that the minimum in the dilaton direction is $\chi_0 = f_{\chi}$ and at zero vacuum energy, we determine $A$ and $V_0$, including the $\mathbb{Z}_2$ breaking term, to be:
\bea
A & = \frac{1}{4} \left( 1 + 2\frac{\delta_1^2}{\delta_2}(\gamma_m-\gamma_{4f})-8 \delta_2 (3-\gamma_{4f})+2(3-\gamma_m) \delta_3 \sqrt{4-\frac{\delta_1^2}{\delta_2^2}}  \right), \\
V_0 & = \frac{\lambda_{\chi} f_{\chi}^4}{16} \left( 1 + 2 \frac{\delta_1^2}{\delta_2} (2+\gamma_m-\gamma_{4f}) - 8 \delta_2 (1-\gamma_{4f}) - 2 (1+\gamma_m) \delta_3 \sqrt{4-\frac{\delta_1^2}{\delta_2^2}} \right).
\eea
As the pion fields behave like angles, the VEV solution for $\delta_3 = 0$ has two degenerate values. The $\delta_3$ coupling generates a small difference in the values and, more importantly, a  vacuum energy density gap between the two vacua:
\bea
\Delta V = V(\phi_0,\chi_0) - V(-\phi_0,\chi_0) = \lambda_{\chi} f_{\chi}^4 \delta_3 \sqrt{4-\frac{\delta_1^2}{\delta_2^2}}.
\eea

\subsection{Theoretical constraints from compositeness consistency}

The underlying  composite model imposes some constraints on the parameters in Eq.~\eqref{eq:potentialV}. This is not an absolute requirement, but rather a reasonable setup suggesting the size of the effective parameters. Terms involving $\delta_{1}$ and $\delta_{3} $ typically originate from a bare mass term of the confining fermions, while $\delta_{2}$ comes from four-fermion interactions. These terms will generally interact with the dilaton, and in this work we consider the case where the scaling dimensions of the fermion operators determine the $\chi$ couplings to these terms. The dilaton couplings to pions, therefore, depend on two  anomalous dimensions $\gamma_{m}$ and $\gamma_{4f}$. Lattice computations of walking theories~\cite{DeGrand:2015zxa} suggest $0<\gamma_{m}\lesssim 1$, however, its exact value may be strongly modified by additional contributions from scalar mesons or the $U(1)$ axial anomaly of the theory \cite{Ishida:2019wkd}. Therefore, in our analysis, we consider two separate regimes:
\begin{equation}
0<\gamma_{m}\,,~\gamma_{4f} \lesssim 1, \qquad \mbox{and} \qquad
1<\gamma_{m}\,,~\gamma_{4f} \lesssim 2 \,.
\end{equation}

Consistency with the underlying compositeness further constrains the theory space. Terms associated with the pion potential should be a small perturbation to the chiral Lagrangian at the dilaton potential's minimum $\chi = f_\chi$, which further implies that these coefficients will be determined through the pion decay constant $f_{\phi}$. Therefore we apply an order-of-magnitude bound on the couplings $\delta_1$, $\delta_2$ and $\delta_3$ as \cite{Cacciapaglia:2023kat} 
\begin{align}
\operatorname{max}(\delta_{1}, \delta_{2}, \delta_{3}) \lesssim \frac{1}{\lambda_{\chi}} \left(\frac{f_{\phi}}{f_{\chi}} \right)^{4}. 
\end{align}
Furthermore, as the chiral symmetry breaking necessarily implies breaking of the scale symmetry, while the inverse is not always true, we can consider $f_\chi \geq f_\phi$ as a consistency condition (see \cite{CruzRojas:2023jhw,CruzRojas:2025qcj} for an holographic result confirming this ansatz).
{The composite nature of the underlying theory therefore gives a reasonable parameter range for the model parameters in the potential. 
Within these conditions we show that the overall potential can lead to large scale observables consistent with CMB observations, and small scale features leading to copious PBH and GW production. 
}

\section{Background evolution}
\label{sec:background}
We first compute the background evolution of the inflaton in the two-field space defined by $\chi$ and one direction in the pion space, $\phi$. The evolution around the Friedmann-Lema{\^i}tre-Robertson-Walker (FLRW) metric
\bea
d s^2=-d t^2+a(t)^2 \delta_{i j} d x^i d x^j
\eea
gives the following equations of motion for $\chi$ and $\phi$, 
\begin{equation}
D_t \dot{\Phi}^a+3 H \dot{\Phi}^a+G^{a b} D_b V  =0,
\end{equation}
and the Friedmann equation,
\begin{equation}
3 H^2  =\frac{1}{2} \dot{\Phi}_0^2+V,
\label{eq:background}
\end{equation}
where $\Phi = (\chi, \phi)^{T} $ and  $\dot{\Phi}_{0}^2 = G_{ab} \dot{\Phi}^{a} \dot{\Phi}^{b} $. Here, $G$ is the metric in field space defined in Eq.~\eqref{eq:Gkin} and $V$ is the potential in Eq.~\eqref{eq:potentialV}, while the covariant derivative in the curved field space is defined as $DA^a\equiv dA^a+\Gamma^{a}_{bc}\dot\phi^cdA^b$, where $\Gamma^{a}_{bc}$ is the Christoffel symbol for the field metric $G$. Accordingly, $D_t\equiv D/dt$, $D_a=D/d\Phi_a$.
The solutions trace a unique trajectory in the field space, which allows us to define a tangential and a normal direction in terms of the following vectors: 
\bea
T^a=\frac{\dot{\Phi}^a}{\dot{\Phi}_0}, \quad N_a=\sqrt{\operatorname{det} G} \epsilon_{a b} T^b.
\eea
Henceforth, the slow-roll parameters $\epsilon$ and $\eta$, which govern the background evolution, can be written as 
\bea \left\{ \begin{array}{l}
\epsilon \equiv-{{\dot{H}}/{H^2}}={\dot{\Phi}_0^2}/{2 H^2} ~, \\
\eta^a=\eta_{\|} T^a+\eta_{\perp} N^a ~, \end{array} \right. \qquad \mbox{with} \quad
\eta_{\|} \equiv-\frac{\ddot{\Phi}_0}{H \dot{\Phi}_0}, \quad \eta_{\perp} \equiv \frac{U_N}{\dot{\Phi}_0 H} .
\eea
Introducing the e-folding number $\mathcal{N}$ as a new time variable, we rewrite the background equations as
\begin{align}
    &\chi''+(3-\epsilon)\chi'+(3-\epsilon)\frac{V_{,\chi}}{V}=\frac{\phi'^2\chi}{f_{\chi}^2},\label{eq:chieom}\\
    &\phi''+(3-\epsilon)\phi'+\left(\frac{f_{\chi}}{\chi}\right)^2(3-\epsilon)\frac{V_{,\phi}}{V}=-2\frac{\phi'\chi'}{\chi},
     \label{eq:phieom}
\end{align}
where $X'=dX/d\mathcal{N}$ and $V_{, x} = D_x V$ is the field derivative of the potential. 
\begin{figure}[t!]
    \centering
    \includegraphics[width=0.8\linewidth]{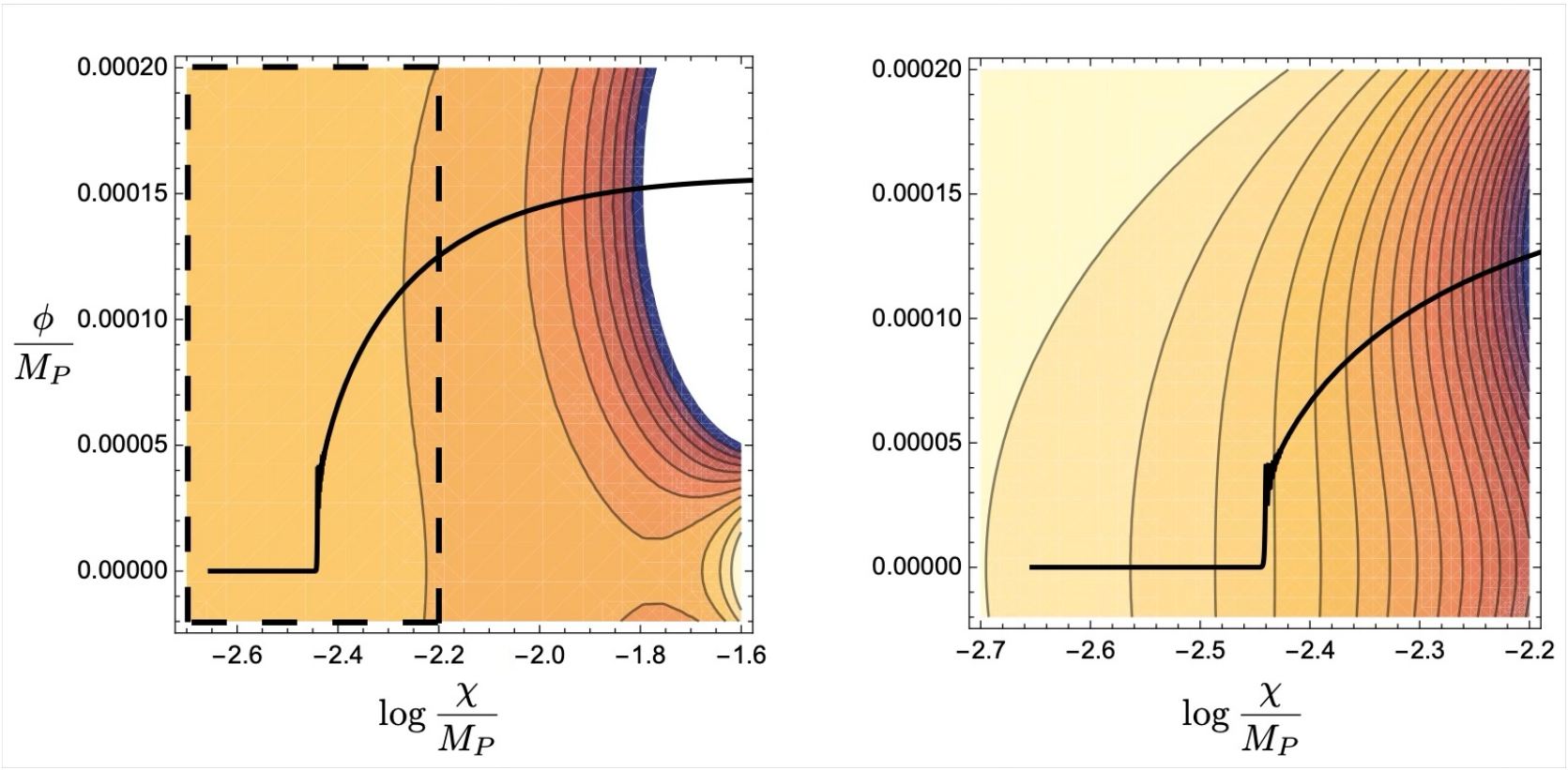}
    \caption{Field trajectory (solid line) in the $(\chi, \phi)$ plane, with the contours representing the potential height. The full trajectory depicts a turn (left), where a zoomed-in figure displays the tachyonic hill responsible for the turn (right).}
    \label{fig:trajectory}
\end{figure}
\begin{figure}[t!]
    \centering
    \includegraphics[width=0.5\linewidth]{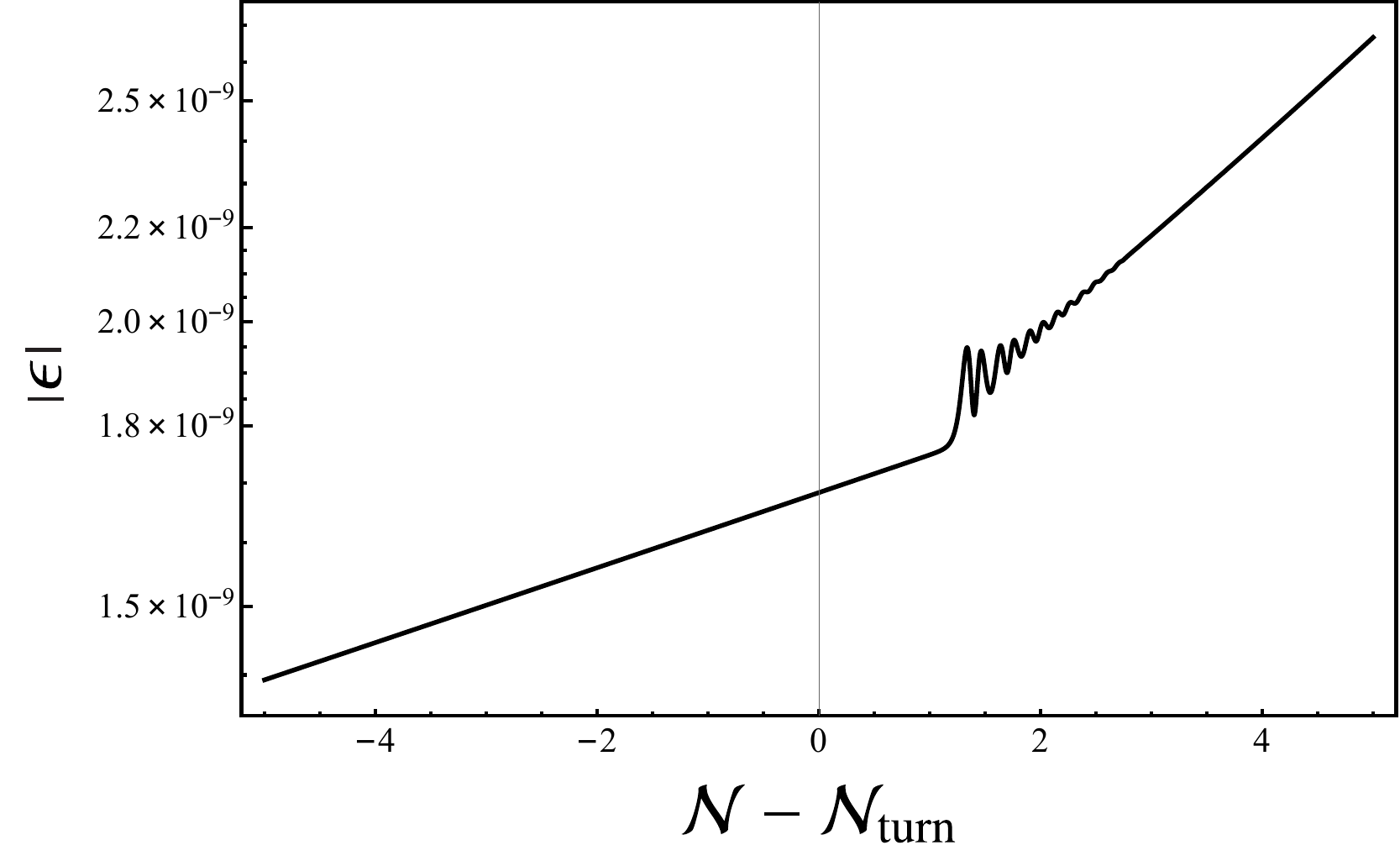} \hspace{0.5cm}
    \includegraphics[width=0.45\linewidth]{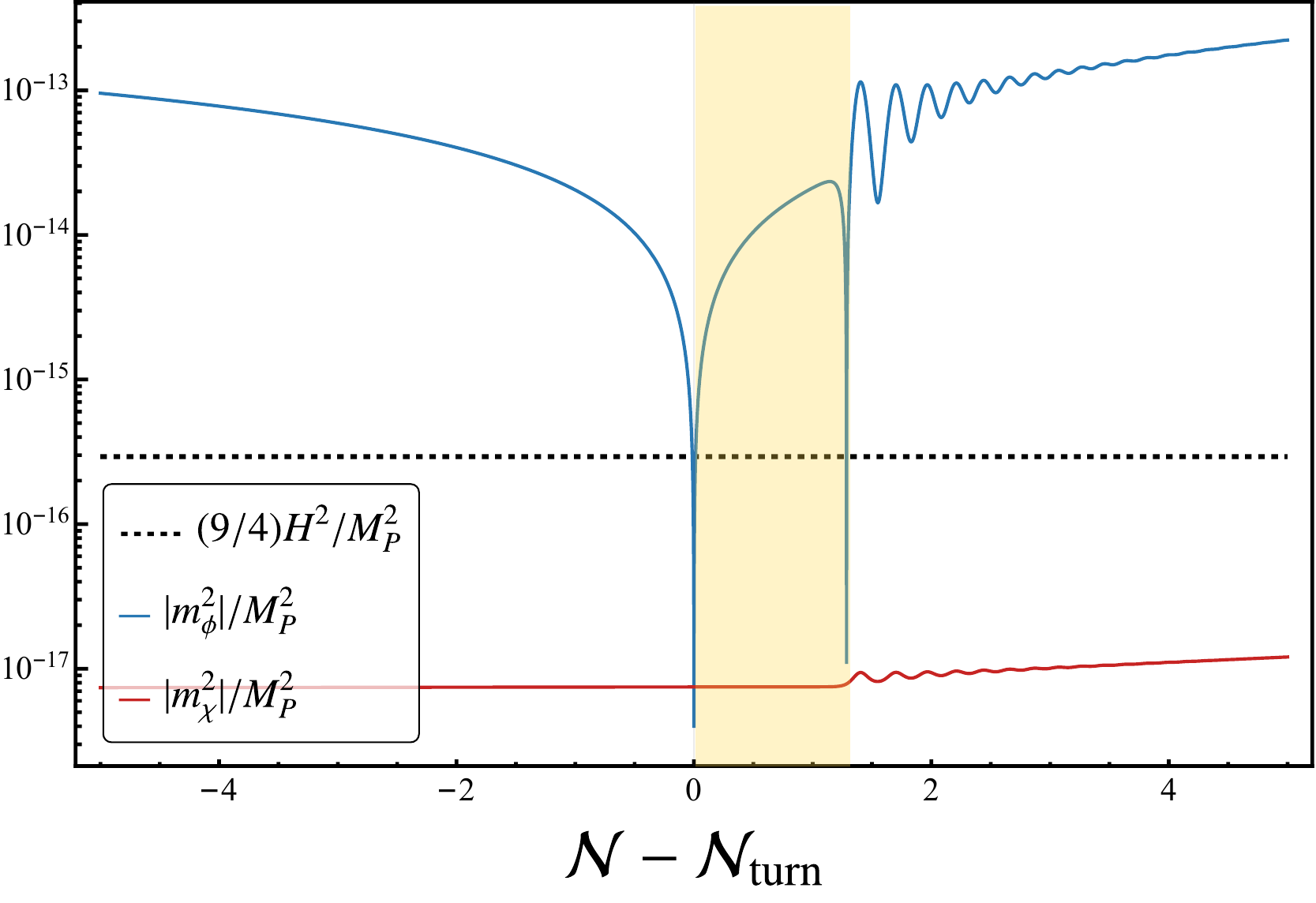}
    \caption{Evolution of $\epsilon$ (left) and of the effective scalar masses (right) as a function of $\mathcal{N}$. For both plots, we solve the background evolution equations \eqref{eq:background} for a specific set of parameters (corresponding to set 2 in Table \ref{tab:Benchmark}). 
     In the right plot, the yellow-shaded region shows the stage that the background is dominated
by a tachyonic field $\phi$, i.e., $m_{\phi}^2< 0$, while the horizontal dotted line indicates the value of the Hubble parameter. 
}
    \label{fig:slowrollparameter}
\end{figure}
Adapting the slow-roll condition to the background equation of motion, $\epsilon\ll1$, we obtain
\begin{align}
    &\chi'=-\frac{V_{,\chi}}{V},\\
    &H^2=\frac{V}{3}\approx\frac{V_0}{3}.
\end{align}

As a reference, we illustrate the inflaton trajectory in Figure~\ref{fig:trajectory}, while in Figure~\ref{fig:slowrollparameter} (left) we plot the slow-roll parameter $\epsilon$, showing that it remains small over the whole trajectory. Initially, starting from $\chi \ll f_\chi$, the inflaton stays in the valley defined by $\phi\simeq0$, where the pion potential is dominated by the $\delta_{1}$ term~\footnote{{$\phi_{i}$ is a heavy field during the inflationary period, therefore quickly resides to the VEV, making the initial value $\phi_{i}$ dependence of observables negligible.}}. Essentially, this defines a first evolution phase corresponding to single-field inflation.  As the $\delta_2$ term becomes more relevant, it passes through a saddle point in the potential, where the inflaton experiences a tachyonic instability leading to an exponential growth of the isocurvature perturbations. The inflaton then falls down into the valley dominated by the $\delta_{2}$ term, mixing the isocurvature and curvature fluctuations, then rolls down in the $\chi$ direction, again into an effective single-field regime. Accordingly, we can divide the inflationary dynamics into 3 stages determined by the effective masses of $\chi$ and $\phi$, i.e. $m_{\chi}^2=V_{,\chi\chi}$ and $m_{\phi}^2=(f_\chi/\chi)^2V_{,\phi\phi}$. As an illustration, we show in Figure~\ref{fig:slowrollparameter} (right) the masses as a function of $\mathcal{N}$. Henceforth, the three stages are defined as:
\begin{itemize}
\item Stage~1 [$m_{\chi}^2<m_{\phi}^2$]: The inflaton rolls stably along the $\phi\simeq0$ potential valley along the $\chi$ direction. The large pion mass is due to the $\delta_1$ term, and  this phase of inflation can be understood as an effective $\chi$-driven slow-roll. This phase terminates when the inflaton reaches the saddle point $(\chi_c, \phi_c)$, which can be obtained through $\left.V_{,\phi\phi}\right|_{\phi=\phi_c,\chi=\chi_c}=~0$ and $V_{,\phi}=0$.  This yields:
\begin{align}
    &\phi_c\approx f_{\phi}\arctan{\delta_3/\delta_1}\simeq f_{\phi}\delta_3/\delta_1\ll f_\phi,\\&\chi_c\simeq f_{\chi}\left(\frac{\delta_1}{2\delta_2}\right)^{\frac{1}{3-2\gamma_{4f}+\gamma_m}}.  \label{eq:chic}
\end{align}
This point was used to terminate inflation in \cite{Cacciapaglia:2023kat}.

\item Stage~2 [$m_{\chi}^2>m_{\phi}^2$]: A tachyonic phase starts beyond the saddle point $(\chi_c,\phi_c)$ due to a negative mass squared for the pion field. 
 Hence, the inflaton trajectory makes a turn in the field space to the $\phi$ direction and start to slow-roll along $\phi$-direction until the minimum of the $\phi$ potential is reached. At the end of Stage~2, the inflaton becomes massive and oscillates around the local potential minimum, being stabilized at $\phi\sim\phi_0$ given by
\begin{align}
    \phi_0(\chi)\simeq\pm f_{\phi}\arccos\left[\frac{f_{\chi}^3\delta_1}{2\delta_2\chi^3}\left(\frac{\chi}{f_\chi}\right)^{2\gamma_{4f}-\gamma_m}\right].
    \label{eq:localmin}
\end{align}
This field configuration provides the initial conditions for the next stage of inflation.
 \item Stage~3 [$m_{\chi}^2<m_{\phi}^2$]: After the pion field settles at the minimum, dominated by the $\delta_2$ term, the dilaton $\chi$ drives inflation once again. This last stage of inflation ends due to the violation of the slow-roll condition near the global minimum of the inflation potential, which is given by $\left.V_{,\chi}\right|_{\phi=\phi_q,\chi=\chi_q}=\left.V_{,\phi}\right|_{\phi=\phi_q,\chi=\chi_q}=0$. 
\end{itemize}

As the slow-roll conditions are only briefly violated during Stage~2, the total duration of inflation can be reasonably estimated follow the slow-roll approximation~\cite{Pi:2017gih,Planck:2018jri,Cheong_2020,Cheong_2022}: 
\begin{align}
N_{\text{tot}} \simeq 61.6-\frac{1}{12} \ln \left(\frac{45 V_0}{\pi^2 g_s T_{\mathrm{reh}}^4}\right)-\ln \left(\frac{V_0^{1 / 4}}{H^*}\right) ,
\end{align}
where $T_{\rm reh}$ is the temperature of reheating, $g_s$ is the number of relativistic degrees of freedom at the end of reheating (typically taken to be $g_s=106.75$), and $H^*$ is the Hubble scale at the pivot scale $k^*=0.05$ Mpc$^{-1}$. For our values of $V_{0}$, this relation restricts $N_{\text{tot}}$ to be around $48$ to $55$ e-folds.

\begin{figure}[t!]
    \centering
    \includegraphics[width=0.48\linewidth]{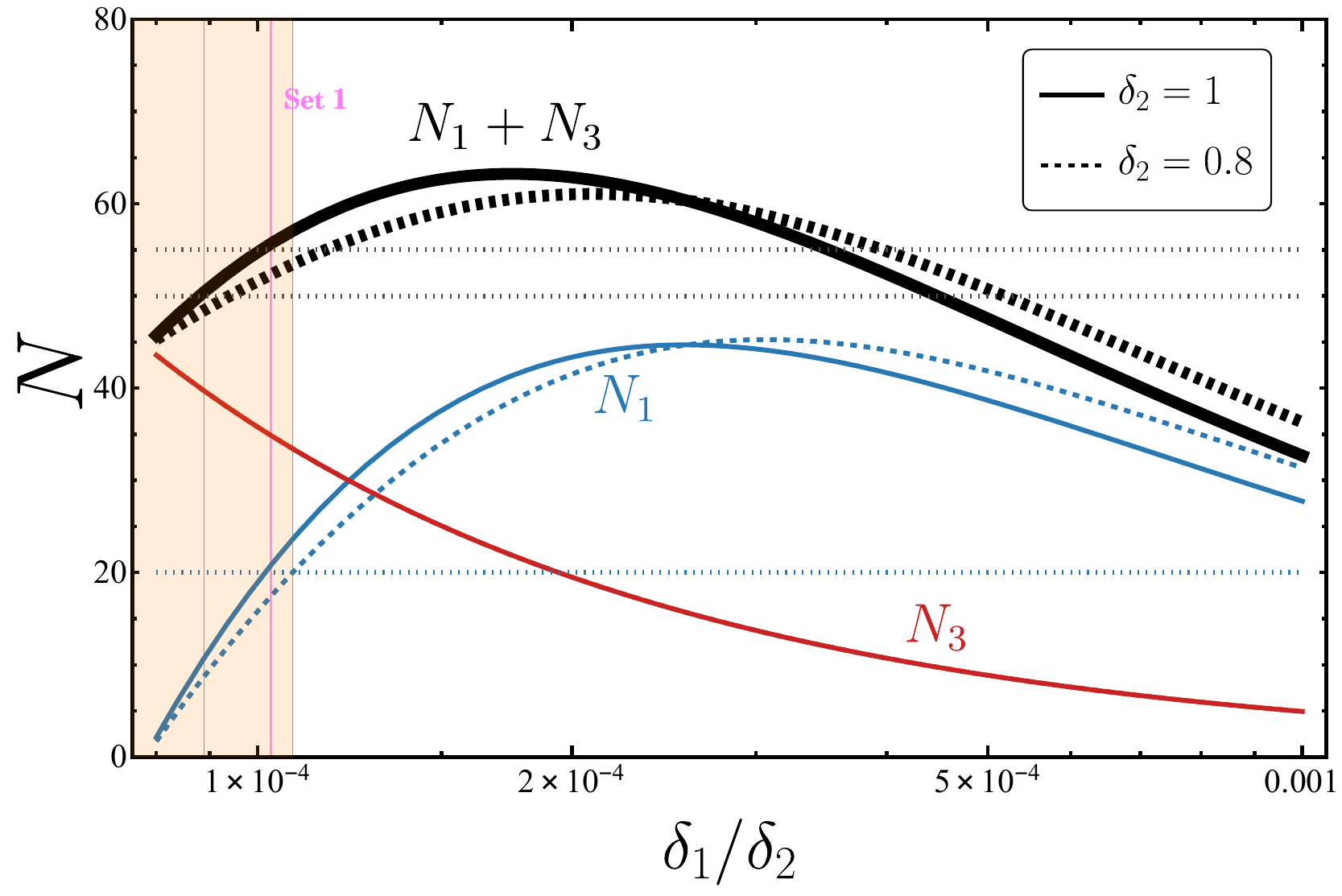}
    \includegraphics[width=0.48\linewidth]{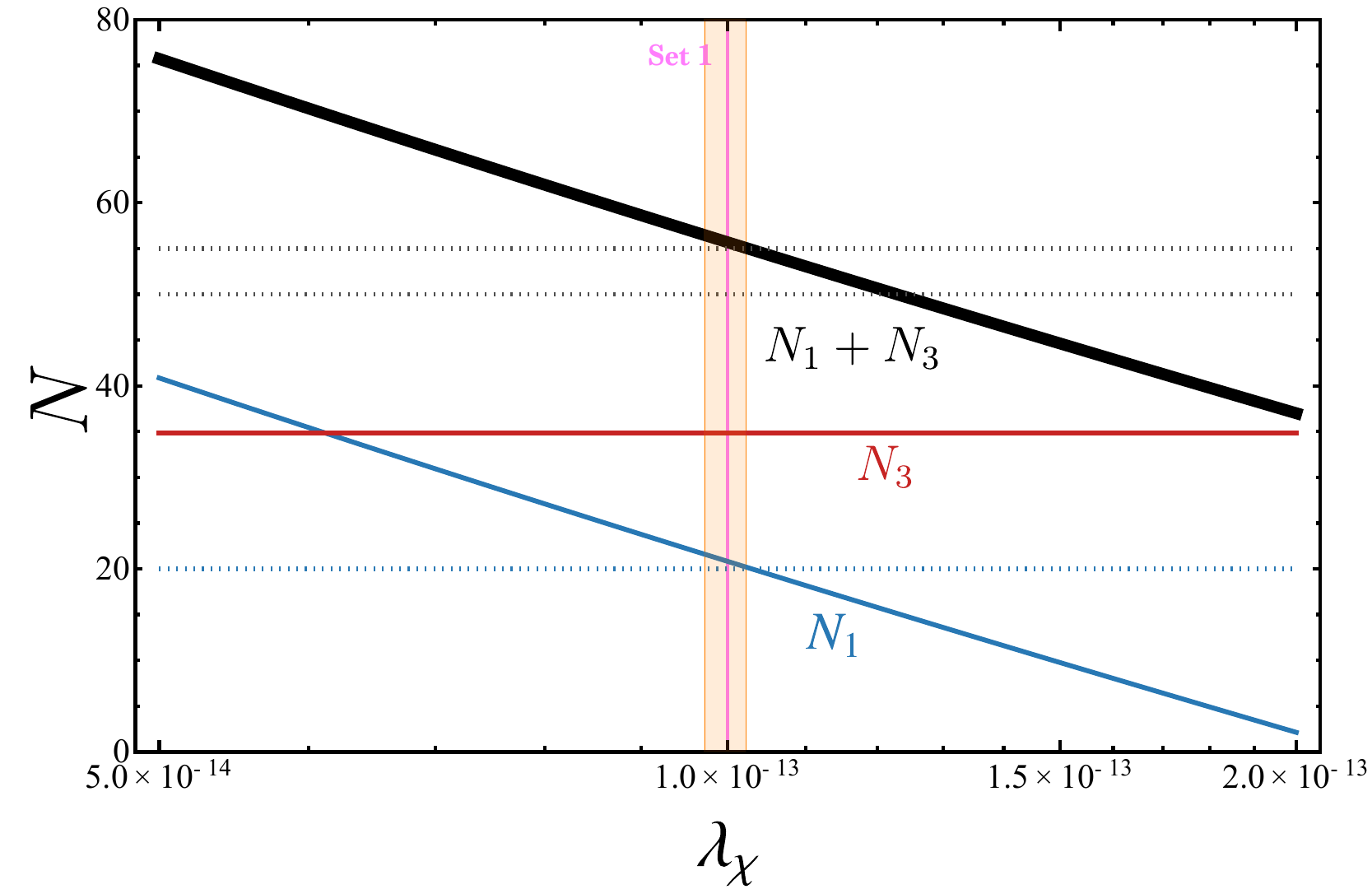}

    \caption{E-folding number $N$ during the stages $1$ and $3$, and the total, as a function of some model parameters with fixed $\gamma_m = \gamma_{4f} = 1$. In the left plot, we use $f_\chi=0.5$, $\lambda_{\chi}=1\times 10^{-13}$ and two values of $\delta_2 = 1$ (solid lines) and $0.8$ (dashed lines). In the right plot, we use $f_\chi=0.5$, $  \delta_{1}=1.03\times 10^{-4}$ and $\delta_{2}=1$. The blue and red lines show the e-folding number of stage 1 and stage 3, while the black line shows the total.  The pink vertical line shows the benchmark set 2 in Table~\ref{tab:Benchmark}, and the green vertical line shows the parameters for $n_s=0.9668$. The orange shaded area shows the parameter space of the model that gives $n_s\in (0.9668 - 0.0037,0.9668 + 0.0037)$ for $\delta_2=1$.}
    \label{fig:efolds}

\end{figure}

The total e-folding number is mainly given by Stage~1 and Stage~3, i.e. the two effective single-field slow-roll stages. For illustration purposes, we provide numerical results in Figure \ref{fig:efolds}, showing the dependence of the e-folding number on $\delta_1/\delta_2$ (left) and on $\lambda_\chi$ (right). Note that we highlighted a benchmark point that will be use in the next section.\,\footnote{Corresponding to set 2 in Table~\ref{tab:Benchmark}.} The numerical curves are derived from semi-analytical formulas we derive below for each stage of inflation.

\subsection*{Stage 1}
The dynamics of Stage~1 and Stage~3 are very similar: both stages exhibit a dilaton-dominated effective single-field inflation accompanied by a heavy auxiliary pion field. 
The pion $\phi$ is stabilized at the potential minimum along the direction $\phi\simeq0$ during Stage~1.
Therefore, $\phi$ hardly contributes to the total e-folding number, and the inflation dynamics are solely determined by $\chi$. In the slow-roll regime, we drop the second time derivative term to solve Eq.~(\ref{eq:chieom}).  Under the slow-roll assumption,  the potential derivative $V_{,\chi}$ at the critical point is given by
\begin{align}
\left.V_{,\chi}\right|_{\phi=\phi_c}&=\frac{1}{4}\lambda_{\chi}\chi^3\left(\mathcal{C}+4\log\frac{\chi}{f_\chi}+4(\gamma_m-3)\delta_1\left(\frac{\chi}{f_\chi}\right)^{-\gamma_m-1}\right)+\mathcal{O}\left(\frac{\delta_3}{\delta_1}\right)^2,
\end{align}
\\where
\begin{align}
    \mathcal{C}=\frac{2}{\delta_2}\left[(\gamma_{4f}-\gamma_m)\delta_1^2+(12-4\gamma_{4f})\delta_2^2\right].
\end{align}
Hence, in order to obtain a turning trajectory, the parameters of the pion potential must satisfy the condition $\delta_1/\delta_2\ll 1$, thus approximately $\mathcal{C}\approx 8(3-\gamma_{4f})\delta_2$.
For $\gamma_{m}=\gamma_{4f}=1$, the total e-folding number of Stage~1 is estimated by the integral
\begin{align}
    N_1\approx-\int_{\chi^*}^{\chi_c}\frac{V_0}{V_{,\chi}}{\rm d}\chi\approx-\frac{f_\chi}{16}\int_{\chi^*}^{\chi_c}\left\{-2\delta_1\left(\frac{\chi}{f_\chi}\right)+\left[4\delta_2+\log\left(\frac{\chi}{f_\chi}\right)\right]\left(\frac{\chi}{f_\chi}\right)^3\right\}^{-1}{\rm d}\chi\,,
    \label{eq:n1}
\end{align}
where $\chi^*=\chi(\mathcal{N}_\text{CMB})$ is the initial condition of the field $\chi$ that satisfies the CMB constraints on the power spectrum $\mathcal{P}_{\mathcal{R}}(\chi^*)=2.1\times 10^{-9}$ and $n_s(\chi^*)=0.9668$.\,\footnote{Here, to avoid confusion, we introduce $N$ to represent the e-folding number within a time interval, while $\mathcal{N}$ represent a time variable.}

\subsection*{Stage 2}
During Stage $2$, the inflation trajectory takes a turn in the field space, rolling in the $\phi$ direction. The phase starts when $\chi$ reaches the saddle point $\chi_c$ in Eq.~\eqref{eq:chic}, 
which satisfies $V_{,\phi\phi}=0$. For  simplicity, we can assume $\phi_{c}\sim 0$ at the start of Stage~2. The end of the Stage~2 is determined by the end of $\phi$ domination, when trajectory is stabilized at the bottom of one of the local potential minima of $\phi$ and restart to slow-roll in $\chi$-direction.
We can then estimate $\phi$ at the end of Stage $2$ by finding the local potential minimum, which gives $\phi_0$ as in Eq.~\eqref{eq:localmin}. 
However, during Stage~2, the tachyonic instability causes the second slow-roll parameter $\eta$ to exceed 1. 
Hence, we cannot estimate the e-folding number of Stage~2 by simply applying the slow-roll assumption. As a solution, we use the WKB approximation to approach the solution. We approximate the time evolution of $\phi$ by
\begin{align}
    \phi(\mathcal{N})=\phi_c\exp(\omega\cdot (\mathcal{N}-\mathcal{N}_c)).
\end{align}
By inserting the WKB solution to the equation of motion of $\phi$, we obtain,
\begin{align}
    \left(\frac{d\omega}{d\mathcal{N}}\right)^2+\frac{d^2\omega}{d\mathcal{N}^2}+3\frac{d\omega}{d\mathcal{N}}+3\left(\frac{f_\chi}{\chi}\right)^2\frac{V_{,\phi}}{\phi V}=0\,,
\end{align}
when taking $\epsilon=0$ and neglecting the $\mathcal{O}(\phi'\chi')$ term. 
Assuming $   \left(\frac{d\omega}{d\mathcal{N}}\right)^2\gg\frac{d^2\omega}{d\mathcal{N}^2} $, we obtain the growing mode of $\phi$
\begin{align}
    \phi'=\phi\frac{d\omega }{d \mathcal{N}}\approx\phi\left(-\frac{3}{2}+\sqrt{\frac{9}{4}-3\left(\frac{f_\chi}{\chi}\right)^2\frac{V_{,\phi}}{\phi V}}\right),
\end{align}
here we neglect the decaying mode since it decays out rapidly.

As a result, we could estimate the total e-folding number of Stage~2 as
\begin{align}
N_2\approx\int_{\phi _c}^{\phi_0}\left[-\frac{3}{2}+\sqrt{\frac{9}{4}-3\left(\frac{f_\chi}{\chi}\right)^2\frac{V_{,\phi}}{\phi V}}\right]^{-1} {\rm d}\phi.
\end{align}

\subsection*{Stage 3}
During Stage~3, the waterfall field $\phi$ is stabilized in one of its potential minima according to the sign of $\delta_3$, and inflation is taken over by $\chi$. Under the slow-roll approximation, when $\phi=\phi_0$, we obtain
\begin{align}
    V_{,\chi}=\lambda_\chi f_\chi^3\frac{\sqrt{4-\left(\frac{\delta_1}{\delta_2}\right)^2\left(\frac{f_\chi}{\chi}\right)^4}\left(\frac{\chi}{f_\chi}\right)^3\log\left(\frac{\chi}{f_\chi}\right)-4\delta_3\left(\frac{\chi}{f_\chi}\right)}{\sqrt{4-\left(\frac{\delta_1}{\delta_2}\right)^2\left(\frac{f_\chi}{\chi}\right)^4}}\approx\lambda_{\chi}\chi^3\log\left(\frac{\chi}{f_\chi}\right),
\end{align}
which goes back to the Coleman-Weinberg (CW) potential by applying the limit of  $\delta_1\ll \delta_2$ and  $\delta_3\ll 1$. The total e-folding number of Stage~3 is therefore approximated by
\begin{align}
N_3\approx-\int_{\chi_c}^{\chi_{f}}\frac{V_0}{V_{,\chi}}{\rm d}\chi\approx\left.-\frac{1}{16}f_\chi^2 \text{Ei}\left[-2\log\left(\frac{\chi}{f_\chi}\right)\right]\right|_{\chi_c}^{\chi_f},
    \label{eq:n3}
\end{align}
where $\chi_f$ is the dilaton field value determined at the end of inflation and $\text{Ei}$ is the exponential integral. Inflation ends when the slow-roll condition is violated, $\epsilon=(\chi')^2/2+2(\phi')^2\chi^2 \simeq(V_{,\chi}/V)^2/2 =1$ in the slow-roll regime. This indicates
\begin{align}
    \chi_f\approx f_\chi \exp\left[\frac{1}{3}W_0\left(-\frac{3 f_\chi}{8\sqrt{2}}\right)\right]\approx f_\chi-\frac{f_\chi^2}{8\sqrt{2}}+\mathcal{O}(f_{\chi}^{3})\,,
\end{align}
where $W_0$ is the principal branch of the Lambert $W$ function. Therefore, $N_3$ is only a function of $\delta_1/\delta_2$ and $f_{\chi}$. When we fix $f_\chi$,  $N_3$ is a monotonically decreasing function of $\delta_1/\delta_2$ as depicted in Figure \ref{fig:efolds}. To satisfy $N_{\text{tot}}\approx N_1+N_3 \sim  50 $ and $N_1 \in (30, 40)$, we need $N_3$ to be in the range $N_{3} \in (10,20)$.

\section{Evolution of the perturbations}
\label{sec:perturbation}

The non-trivial trajectory of the inflaton through the potential gives rise to enhancements in the isocurvature perturbations, and consequently in the curvature one. Working in the spatially flat gauge, with 
\begin{align}
    \Phi^a(t, \vec{x}) &= \Phi_0^a (t) + \delta \Phi^a (t,\vec{x}), \\
    ds^2 &= -(1+2 \psi) dt^2 + a(t)^2(1-2\psi)\delta_{ij} dx^i dx^j \,,
\end{align}
the comoving curvature and isocurvature perturbations are defined as~\cite{GrootNibbelink:2001qt, DiMarco:2002eb, Peterson:2011yt, Greenwood:2012aj,  Cespedes:2012hu, Achucarro:2012yr, Kaiser:2013sna} 
\begin{align}
    \mathcal{R}  & \equiv \frac{H}{\dot{\Phi}_0} Q_T \,,\\
    \mathcal{S}  & \equiv \frac{H}{\dot{\Phi}_0} Q_N .
\end{align}
The Mukhanov-Sasaki variables  $Q^a \equiv \delta \Phi^a + \frac{\dot{\Phi}^a}{H}\psi$ follow the perturbation equations:
\begin{align}
    &\ddot{\mathcal{R}} + \left(3 + 2\epsilon - 2 \eta_{\parallel} \right)H \dot{\mathcal{R}} + \frac{k^2}{a^2} \mathcal{R} = -2 \frac{H^2}{\dot{\Phi}_0} \eta_\perp \left[\dot{Q}_N + \left(3 - \eta_\parallel + \frac{\dot{\eta}_\perp}{H \eta_\perp}\right) H Q_N  \right] \,,
    \label{eq:adiabaticperteom}\\
    &\ddot{Q}_N + 3 H \dot{Q}_N + \left(\frac{k^2}{a^2}+ M_\text{eff}^2 \right)Q_N = 2 \dot{\Phi}_0 \eta_\perp \dot{\mathcal{R}} \,,
    \label{eq:isocurvatureperteom}
\end{align}
where the isocurvatue perturbation $Q_{N}$ comes with an effective mass term $M_\text{eff}^2$ 
\begin{align}
    M_\text{eff}^2  = V_{NN} + H^2 \epsilon \mathbb{R} - \dot{\theta}^2 .
\label{eq:isocurvaturemass}
\end{align}
Here, $V_{NN}$ is the second derivative of the potential with respect to the isocurvature direction, $\epsilon$ is the slow-roll parameter, and $\mathbb{R}$ is the curvature of the field space metric. The term $\dot{\theta}^2$ represents the turn rate of the tangential direction $T_{a}$, with the expression $\dot{\theta} = H \eta_{\perp}$. Depending on the magnitude, $M_\text{eff}^2$ either induces an exponential growth or a damping of the perturbations, as one can deduce from Eq.~\eqref{eq:isocurvatureperteom} and Eq.~\eqref{eq:isocurvaturemass}. The dynamics further mix $Q_{N}$ and $\mathcal{R}$, transferring the growth of the isocurvature perturbations into the curvature perturbations. We analyze the perturbation characteristics for each stage of the inflationary evolution. 



\begin{figure}[t!]
    \centering
    \includegraphics[width=0.45\linewidth]{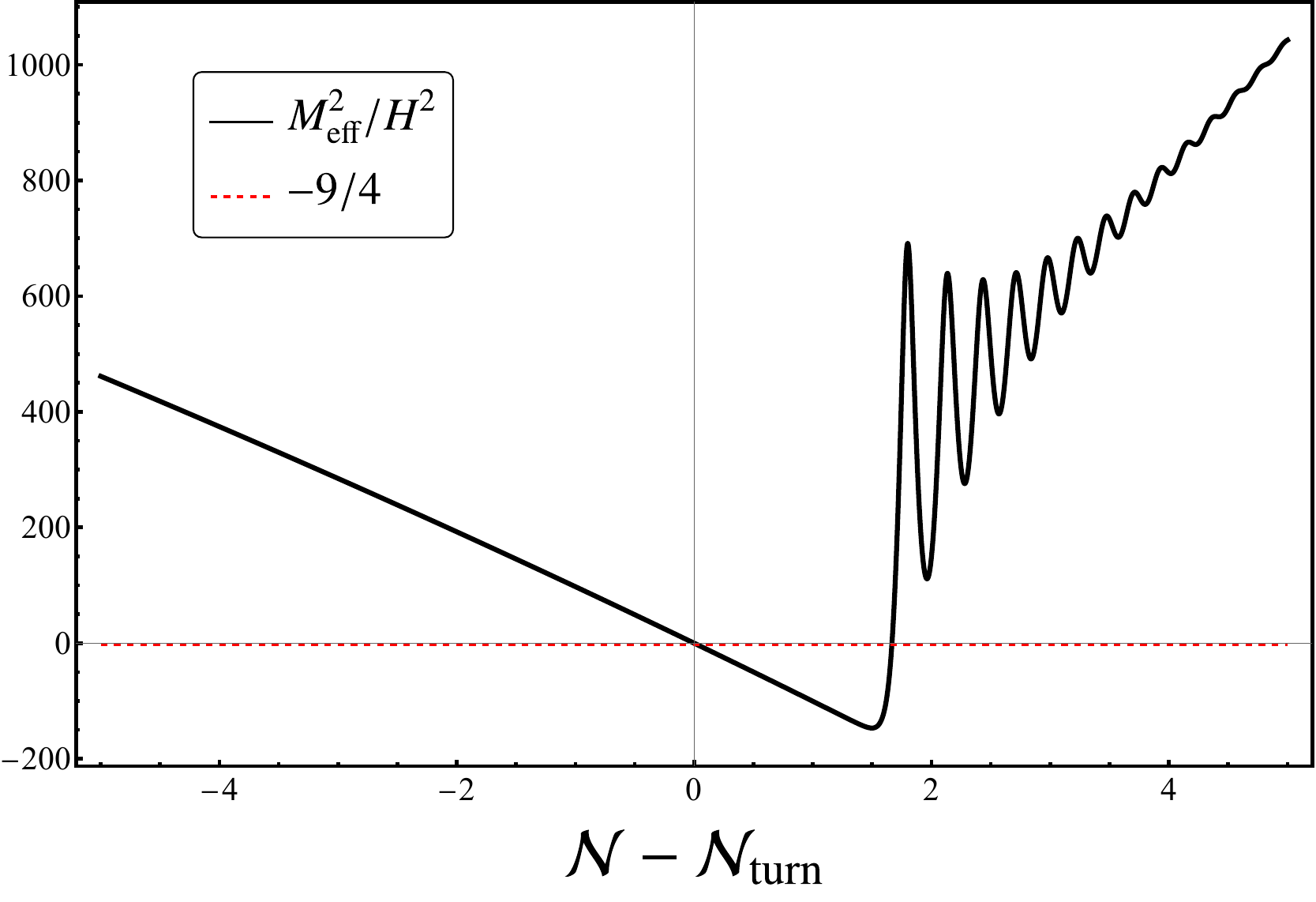}
    \includegraphics[width=0.45\linewidth]{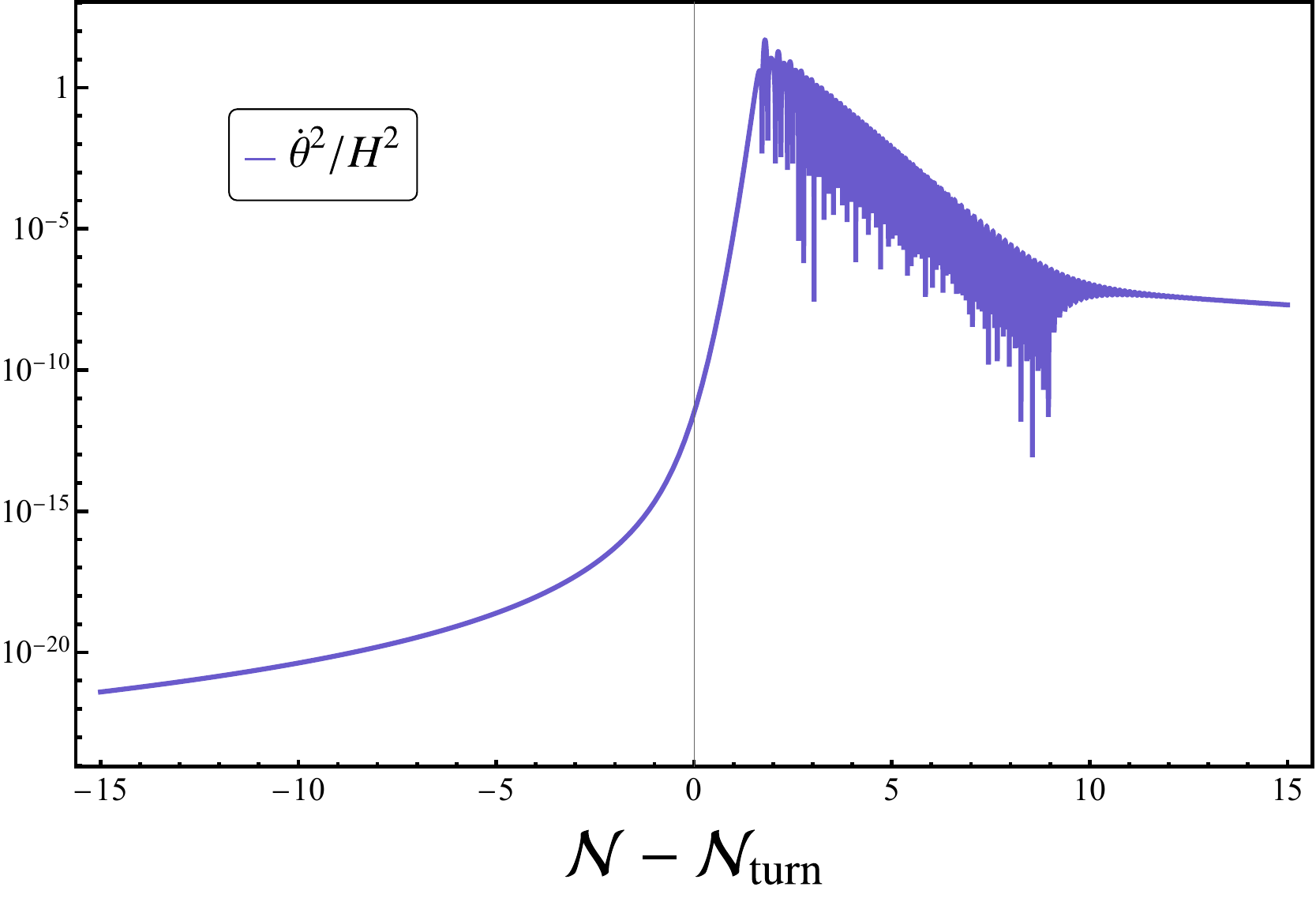}
    \caption{The $M_{\text{eff}}^{2} / H^2$ (left) and the $\dot{\theta}^2 / H^2$ (right) for set 2 in Table~\ref{tab:Benchmark}. The temporary tachyonic stage where $|M_{\text{eff}}^{2}| \gg 9/4 H^2$ induces the isocurvature perturbation growth, and the turn rate $\dot{\theta}^2 \gtrsim H^2$ sources these enhanced isocurvature perturbations to the curvature perturbation.  }
    \label{fig:meffsq_turnratesq}
\end{figure}

{The $M_{\text{eff}}^2 / H^2 $ and $ \dot{\theta}^2  / H^2 $ are depicted in Figure~\ref{fig:meffsq_turnratesq} for a specific benchmark point with $\gamma_m = \gamma_{4f} = 1$.\,\footnote{Corresponding to set 2 in Table~\ref{tab:Benchmark}.} The tachyonic hill in the $\phi$ direction leads to a temporary phase of $M_{\mathrm{eff}}^2 <0$, enhancing the isocurvature perturbations. This enhancement gets transferred to the curvature perturbation due to the period where $\dot{\theta} > H$. Once the inflaton settles back into the minimum, $M_{\mathrm{eff}}^2 >0$ and again exhibits a suppressed isocurvature perturbation. }

{An illustration of the behavior of perturbations during inflation with the same benchmarks is given in Figures~\ref{fig:lwm}, \ref{fig:mwm} and \ref{fig:nwm}. Below we provide more details on the special features of each stage of inflation.}

    
\subsection*{Stage 1}
As previously discussed, the effective single-field inflation dynamics obeys the slow-roll conditions, along with the isocurvature mass $M_\text{eff}^2\simeq m_{\phi}^2  \gg H $. 
The pions $\phi$, therefore, act as massive scalar fields in de Sitter space, with corresponding perturbations being exponentially suppressed. The curvature perturbation power spectrum $\mathcal{P}_\mathcal{R}$ is solely determined by the $\chi$ direction. For $\gamma_{4f}=\gamma_m=1$, $\mathcal{P}_\mathcal{R}$ evaluated at the end of inflation is given by 
\begin{align}
    \mathcal{P}_{\mathcal{R}}^{\mathrm{st-1}}&=\frac{H^2}{4\pi^2} \left(\frac{\partial \mathcal{N}}{\partial \chi }\right)^2 \approx\frac{\lambda_\chi f_{\chi}^{6}}{49152\pi^2(\chi/f_\chi)^{2}}\left[-2\delta_1+\left(4\delta_2+\log\left(\frac{\chi}{f_{\chi}}\right)\right)\left(\frac{\chi}{f_\chi}\right)^{2}\right]^{-2}\,.
\end{align}
The spectral index $n_s$ in this regime will also follow the standard slow-roll expressions, 
\begin{align}
n_{s} ^{\mathrm{st-1}}&\simeq 1 + \frac{d \ln \mathcal{P}_{\mathcal{R}}}{d \ln k} \approx 1+\frac{32}{f_{\chi}^2}\left[-4\delta_1+\left(1+12\delta_2+6\log\left(\frac{\chi}{f_{\chi}}\right)\right)\left(\frac{\chi}{f\chi}\right)^2+\mathcal{O}\left(\frac{\chi}{f_\chi}\right)^4\right].
 \label{eq:cmbob}
\end{align}

\begin{figure}[t!]
    \centering
    \includegraphics[width=0.45\linewidth]{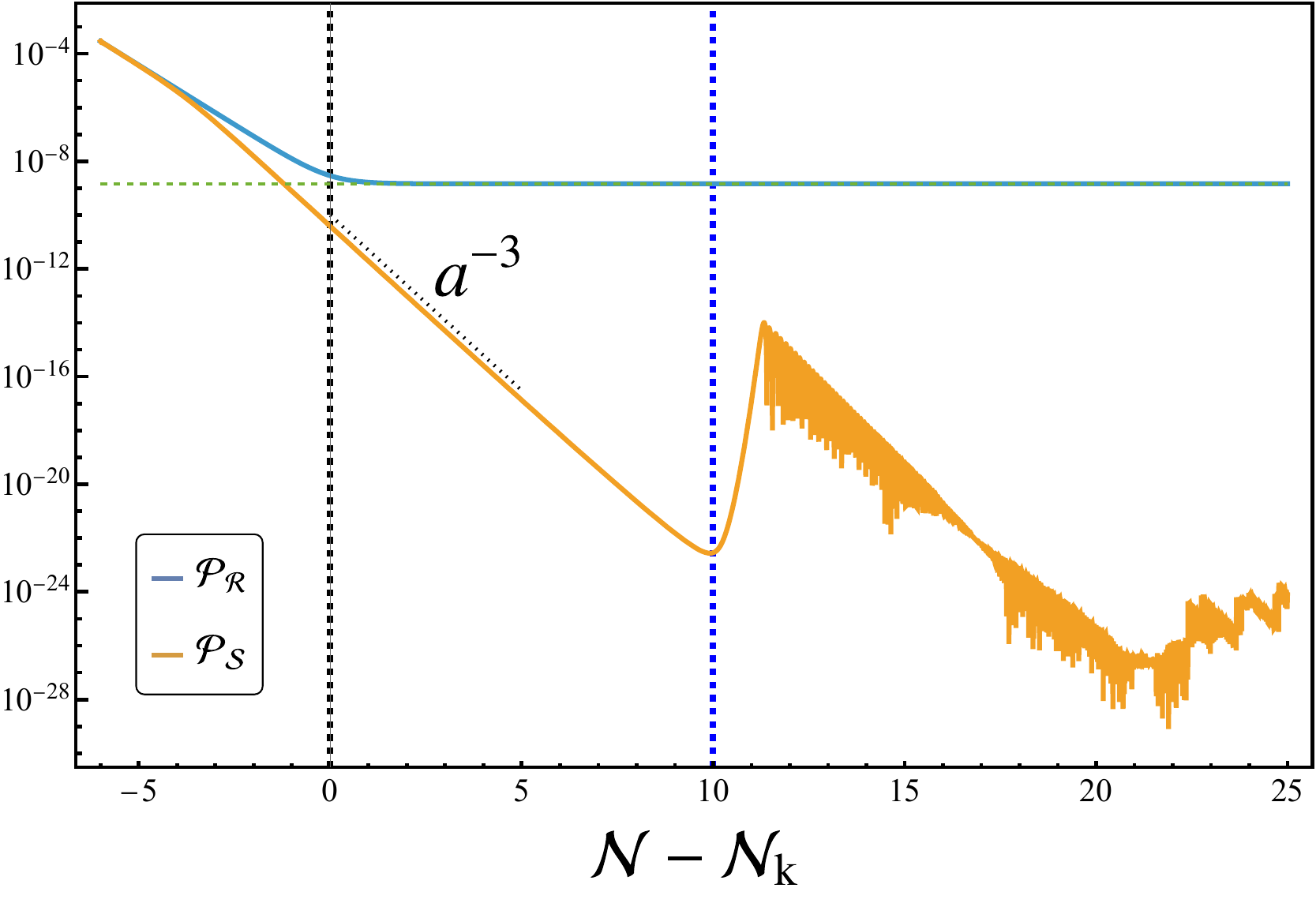}
    \includegraphics[width=0.45\linewidth]{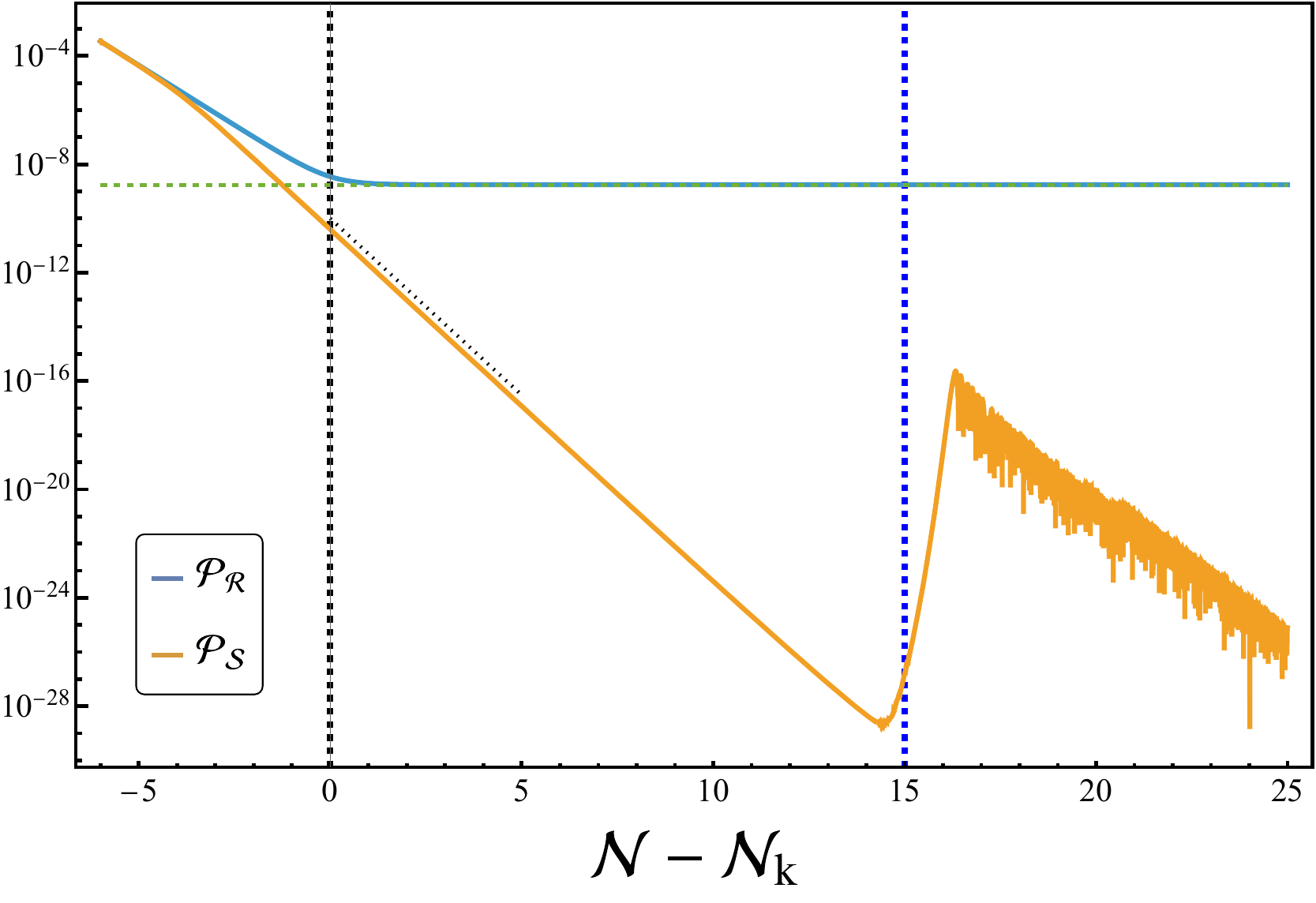}
    \caption{The curvature power spectrum $\mathcal{P}_{\mathcal{R}}(k, \mathcal{N})$(blue) and the isocurvature power spectrum $\mathcal{P}_{\mathcal{S}}(k, \mathcal{N})$(orange) for modes $k = e^{-10}k_{\rm inst}$(left) and $k = e^{-15}k_{\rm inst}$(right), where $k_{\rm inst}\equiv e^{\mathcal{N}_{\rm inst}}H(\mathcal{N}_{\rm inst})$. The black dotted vertical line depicts the horizon exit e-folding number $\mathcal{N}_{k}$ , where the blue dotted vertical line corresponds to the start of the tachyonic growth. For the plot, we numerically solve for the perturbations referring to the Transport method \cite{Dias:2015rca} using the parameters set 2 in Table \ref{tab:Benchmark}.}
    \label{fig:lwm}
\end{figure}

Figure~\ref{fig:lwm} shows the evolution of $\mathcal{P}_{\mathcal{R}}$ and $\mathcal{P}_{\mathcal{S}}$ for a perturbation mode $k \gg k_{c} $ until the end of inflation.\footnote{The wiggling behavior for the $\mathcal{P}_\mathcal{S}$ in the late stage (large $\mathcal{N}$ regime) primarily originates from machine precision. Since at this stage the isocurvature is strongly suppressed and does not affect the evolution of the adiabatic curvature, the numerical result of the final comoving curvature power spectrum $\mathcal{P}_{\mathcal{R}}$ is robust.} We see that,  while the modes are sub-horizon, both $\mathcal{P}_{\mathcal{R}}$ and $\mathcal{P}_{\mathcal{S}}$ exponentially decay in time, where for the latter the positive $M_{\text{eff}}^2$ term governs the decay. Once the perturbations exit the horizon, $ \mathcal{P}_\mathcal{R} $ becomes constant, while the isocurvature perturbations continue to decay. Once the inflaton field passes through the saddle point, the effective mass square $M_\text{eff}^2$ becomes negative, leading to an exponential growth of the isocurvature perturbations. However, this happens when the perturbations are deep outside of the horizon such that $\mathcal{P}_{\mathcal{S}}$ decayed enough prior to this enhancement and $\mathcal{P}_{\mathcal{R}}$ remains constant.

\subsection*{Stage 2}
During this stage, the inflaton passes through the tachyonic $M_{\text{eff}}^2$ region as in Figure~\ref{fig:meffsq_turnratesq}, leading to an exponential growth of the isocurvature perturbations. The effective mass is initially dominated by $m_{\phi}^2 $. Hence, $\mathcal{P}_{\mathcal{S}} (k, \mathcal{N})$ follows the scaling 
\begin{align}
    \mathcal{P}_{\mathcal{S}}(k, \mathcal{N}) \propto \exp \left[\left(\frac{2 |M_\text{eff}|}{H}-3\right) \mathcal{N}\right] .
    \label{eq:scaling1}
\end{align}
Once the inflaton field starts to roll down from the tachyonic hill, $\mathcal{P}_{\mathcal{S}}$ sources $\mathcal{P}_{\mathcal{R}}$, {due to the turn rate exceeding the Hubble rate $\dot{\theta} > H$},  transferring the growth of the isocurvature perturbations into the curvature perturbations, which eventually asymptotes to a constant value deep outside the horizon. 

\begin{figure}[t!]
    \centering
    \includegraphics[width=0.45\linewidth]{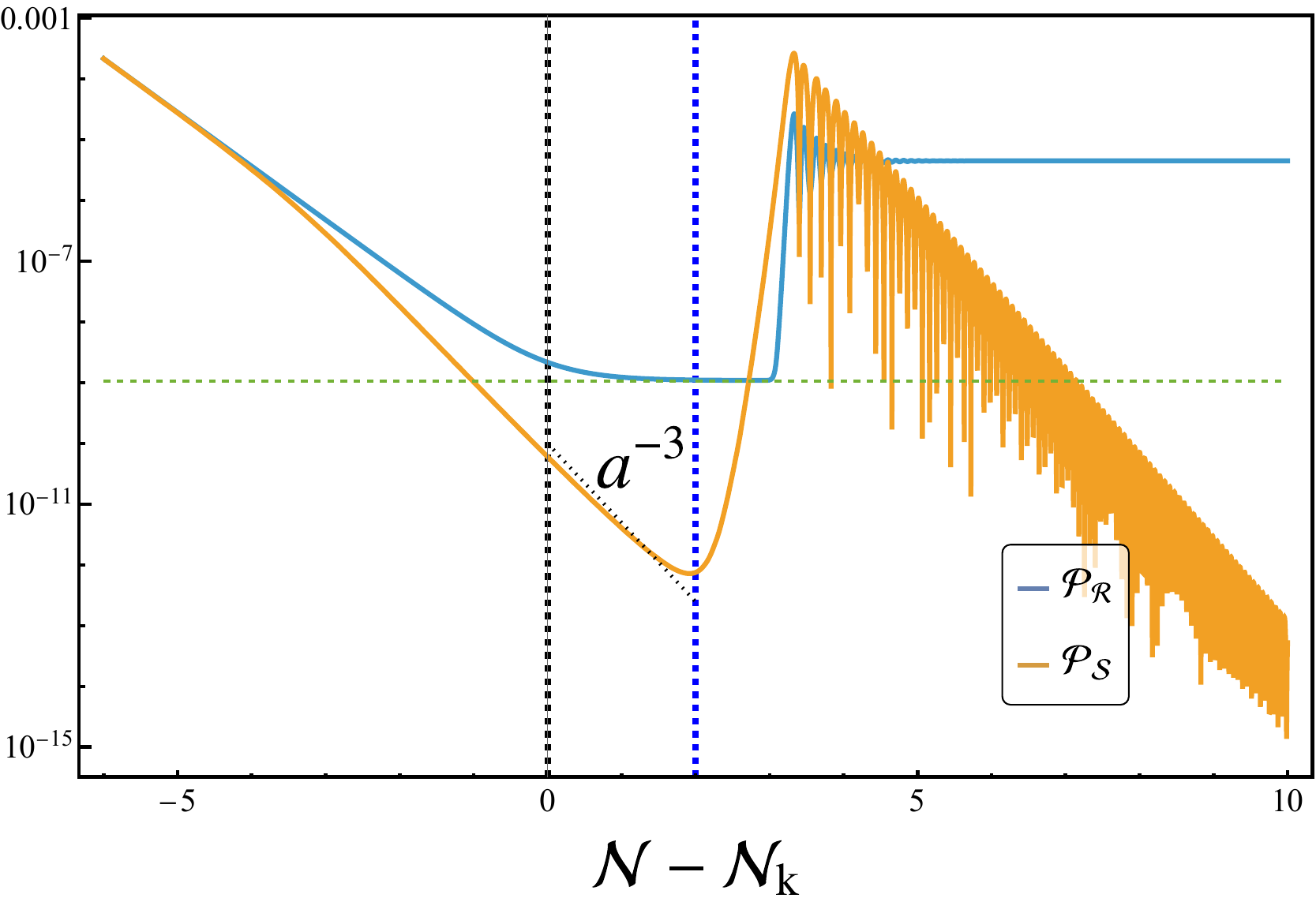}
    \includegraphics[width=0.45\linewidth]{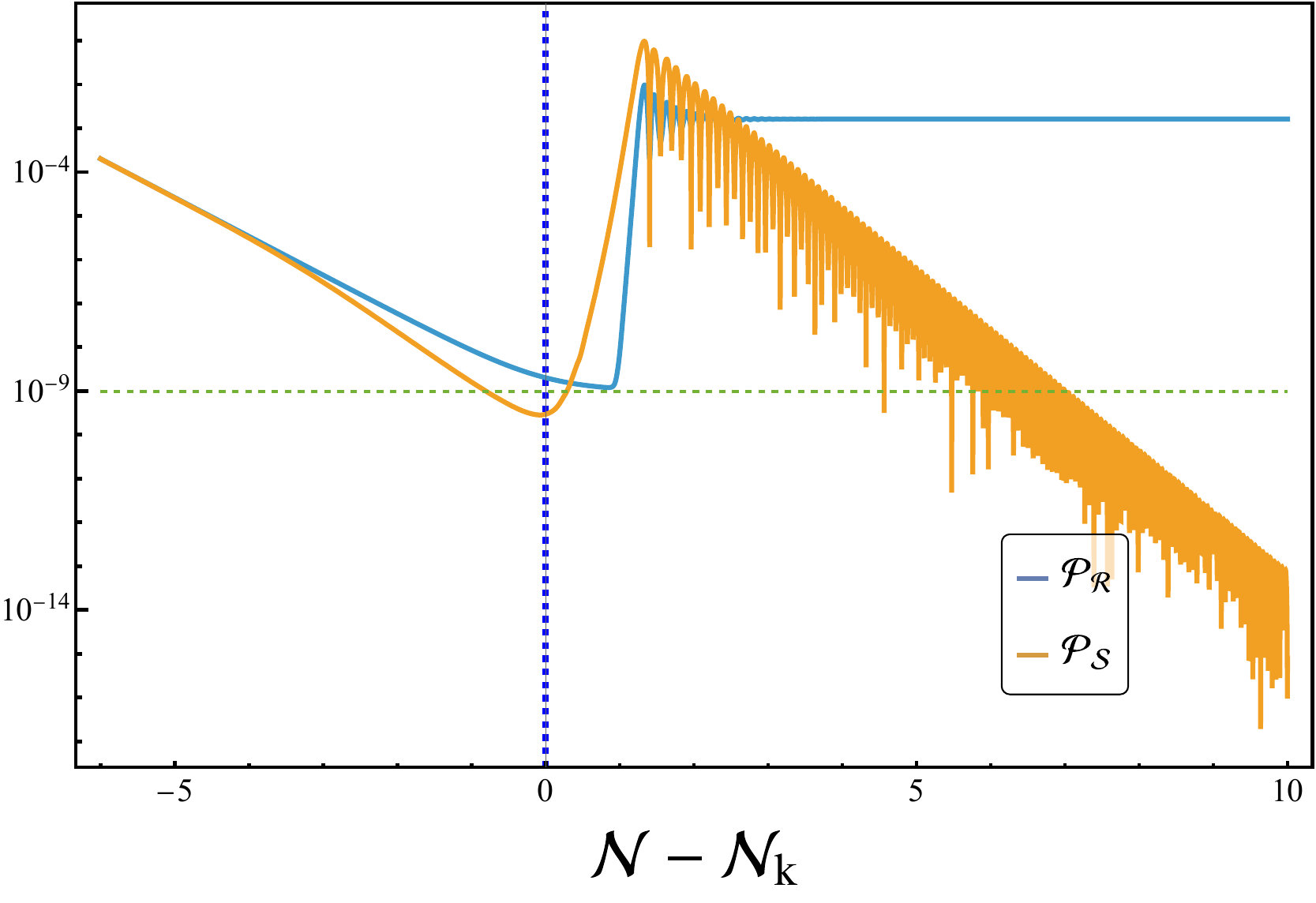}
    \caption{Same plotting scheme as Fig.~\ref{fig:lwm}, but for a mode $k = e^{-2}k_{\text{inst.}} $ (left), where the tachyonic instability occurs after horizon exit, and $k =k_{\text{inst.}}$ (right) where the instability occurs at the horizon exit scale. Both panels show $\mathcal{P}_{\mathcal{S}}$ growing with the relation Eq.~(\ref{eq:scaling1}) to values exceeding $\mathcal{P}_{\mathcal{R}}$, then sourcing it. Once the tachyonic period ends, the isocurvature perturbations decay, while the curvature perturbations remain constant, overall having an enhanced amplitude at the end of inflation.  For the plot, we numerically solve for the perturbations referring to the Transport method \cite{Dias:2015rca} using the parameters set 2 in Table \ref{tab:Benchmark}.}
    \label{fig:mwm}
\end{figure}

\begin{figure}[t!]
    \centering
    \includegraphics[width=0.5\linewidth]{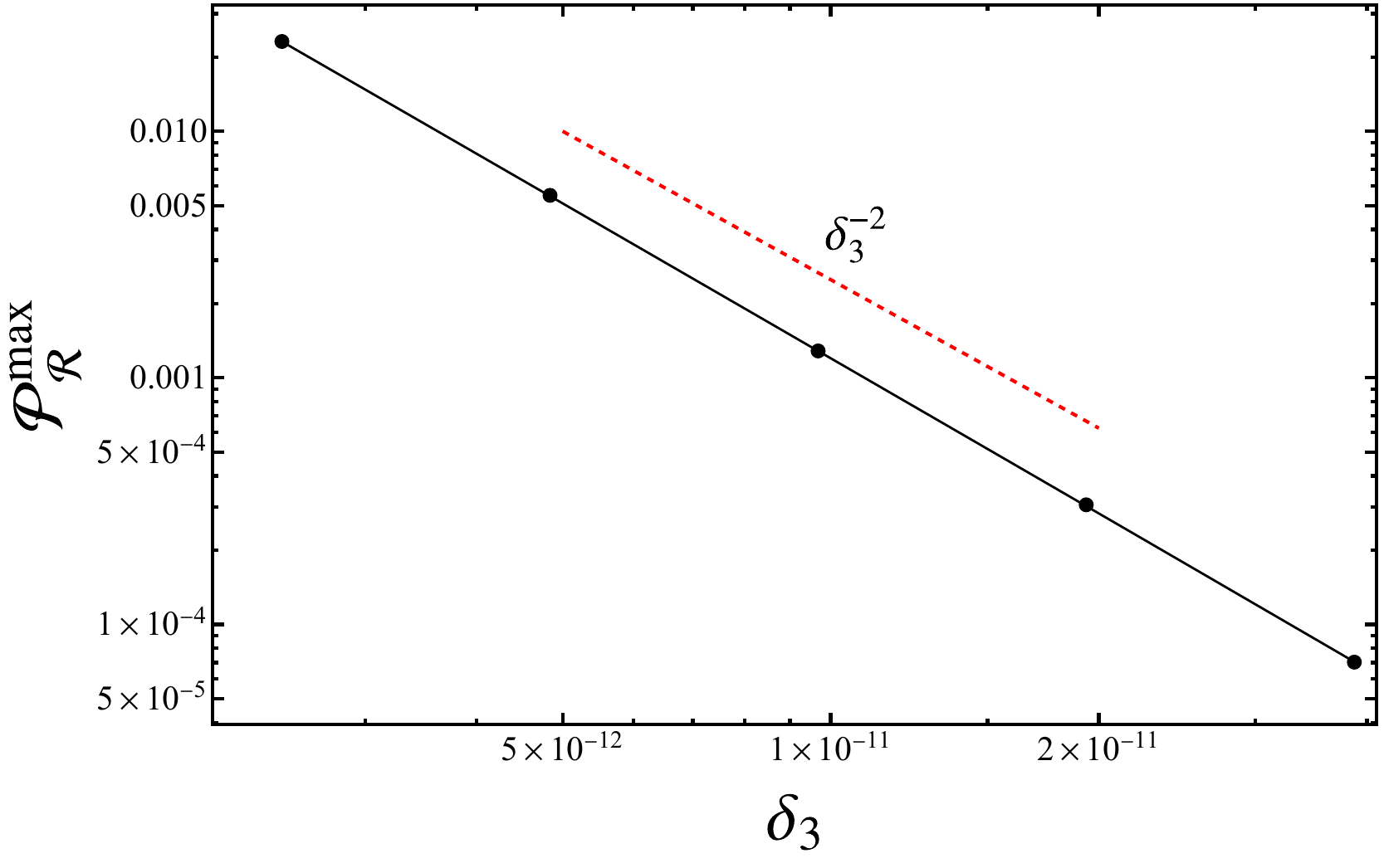}
    \caption{The $\delta_3$ dependence of the peak of the power spectrum. For the plot, we numerically solve for the perturbations referring to the Transport method \cite{Dias:2015rca} using the parameters set 2 in Table \ref{tab:Benchmark} except for a varying $\delta_3$.We found a $\delta_3^{-2}$ dependence as we expected in \eqref{eq:scaling2}.}
    \label{fig:scaling}
\end{figure}

Figure~\ref{fig:mwm} illustrates this behavior for two different $k$ modes. The first mode corresponds to the case where the $\mathcal{N}_{k} < \mathcal{N}_{\text{inst.}}$, with $\mathcal{N}_{\text{inst.}}$ corresponding to the e-folding number at the start of the tachyonic instability, thus $\mathcal{P}_{\mathcal{R}}$ is constant before $\mathcal{P}_{\mathcal{S}}$ induces a growth. The second mode, where $\mathcal{N}_{k} = \mathcal{N}_{\text{inst}}$, shows the enhancement of the curvature perturbations before it asymptotes to a constant, roughly coinciding with the maximum value in $\mathcal{P}_{\mathcal{R}}(k, \mathcal{N}_{\text{end}})$
\begin{align}
    \mathcal{P}_\mathcal{R}^{\max}\approx\left.\left(\frac{\partial N_2}{\partial \phi}\right)^2\right|_{\phi=\phi_c,\chi=\chi_c}\langle\delta\phi^2\rangle
    \propto\frac{48\delta_3}{\phi_c^3f_\phi}\propto \delta_3^{-2}.
    \label{eq:scaling2}
\end{align}
{We obtain the value of $\phi_c$ by solving $V_{,\phi\phi}=0$ when setting $\chi=\chi_c$, which gives $\phi_c\propto\delta_3$. As a result, we observe $\mathcal{P}_\mathcal{R}^{\max}\propto\delta_3^{-2}$ also in the numerical results, see Figure \ref{fig:scaling}. This is consistent with our choice of $\delta_{3}$ as the `tuning' parameter for $\mathcal{P}_{\mathcal{R}}$. Once other couplings, which are associated with $\mathbb{Z}_{2}$ symmetric terms, are fixed by $N_{\mathrm{tot}}$ and $n_{s}$, the $\cancel{\mathbb{Z}}_{2}$ parameter $\delta_{3} $ controls the overall height of $\mathcal{P}_\mathcal{R}^{\max}$ during this stage, determining the quantitative peak size of the small scale $\mathcal{P}_{\mathcal{R}}(k)$ enhancement. }  

\subsection*{Stage 3}

\begin{figure}[t!]
    \centering
    \includegraphics[width=0.5\linewidth]{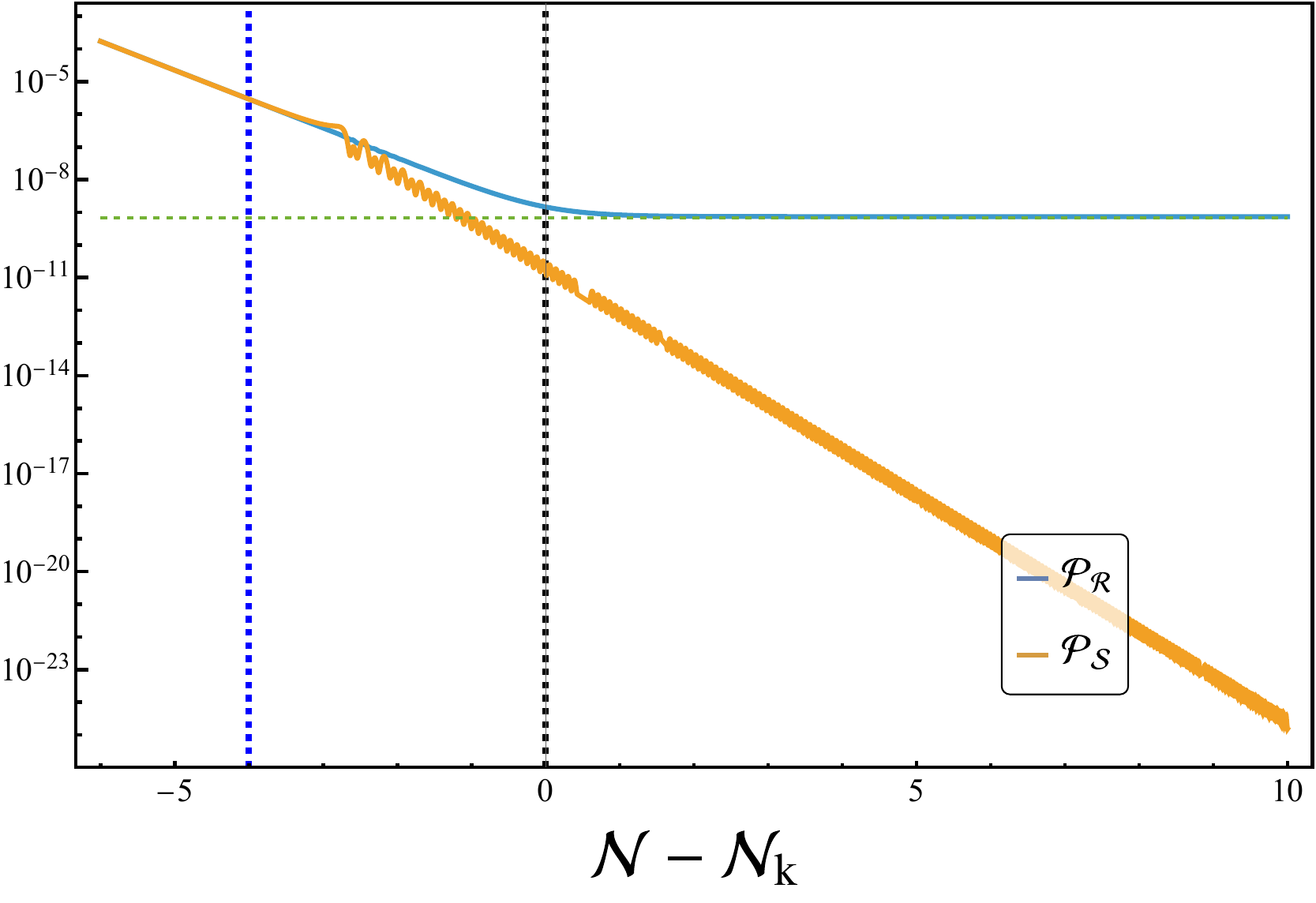}
    \caption{Same plot scheme as Fig.~\ref{fig:lwm} with a mode $k = e^{4}k_{\text{inst.}}$, which enters the horizon after the tachyonic instability stage. The evolutions follow an effective single-field slow-roll phase.  For the plot, we numerically solve for the perturbations referring to the Transport method \cite{Dias:2015rca} using the parameters set 2 in Table \ref{tab:Benchmark}.}
    \label{fig:nwm}
\end{figure}

During the last stage, the dilaton $\chi$ dominates inflation again, leading to another phase of effective single-field inflation. For modes exiting the horizon during Stage 3, the effective mass $M_\text{eff}^2 >0 $  again, leading to a decay of the isocurvature perturbations, resembling the behavior of the first stage.  If the instability occurs deep inside the horizon, $\mathcal{P}_{\mathcal{R}}$ decays again once the instability stage ends, and then the final amplitude will be equivalent to the effective single-field slow-roll inflation. This behavior is illustrated in Figure~\ref{fig:nwm}.

\section{PBH abundance and Stochastic GW background}
\label{sec:gamma1}
The enhanced comoving curvature perturbation $\mathcal{R}$ leads to copious PBH production and second-order stochastic GWs. The PBH mass $M_{\text{PBH}}$ is given by~\cite{10.1093/mnras/168.2.399}
\begin{equation}
\begin{split}
    M_{\text{PBH}} & \simeq 4.64 \times 10^{15} \gamma\left(\frac{g_*}{106.75}\right)^{-\frac{1}{6}}\left(\frac{k_{\mathrm{PBH}}}{k_*}\right)^{-2} M_{\odot} \\
    & \simeq 4.64 \times 10^{15} \gamma\left(\frac{g_*}{106.75}\right)^{-\frac{1}{6}}\exp(-2N_{\text{PBH}}) M_{\odot} \,,
\end{split}
\end{equation}
where $\gamma = 0.2 $ is a $\mathcal{O}(1)$ numerical factor, $g_{*}$ is the number of relativistic degrees-of-freedom, and $N_{\text{PBH}}=\mathcal{N}_{\text{PBH}}-\mathcal{N}_{\text{CMB}}$ is the number of e-folds between the horizon exit $k_{\text{PBH}}$ and the CMB pivot scale $k_{*}$ .
To estimate the PBH abundance, we apply the Press-Schechter formalism \cite{Press:1973iz,Green:2004wb}.
During the radiation domination, assuming linear and Gaussian scalar perturbations, the power spectrum of density contrast $\mathcal{P}_\delta$ is given by
\begin{align}\label{eq:Pw}
    \mathcal{P}_{\delta}(k)=\left(\frac{4}{9}\right)^2(kR)^4\mathcal{P}_{\mathcal{R}}(k)\,,
\end{align}
where the $R=(aH)^{-1}$ is the comoving scale of the Hubble horizon.
The PBH abundance at the time of formation, therefore, can be estimated by integrating the probability distribution function for the collapse of a Hubble patch, as derived from the density contrast threshold $\delta_{\mathrm{th}}$. This collapse probability estimates the probability of formation of a PBH, where $\delta_{\mathrm{th}}=0.3$ is used for the numerical calculation \cite{Carr:1975qj}:
\begin{align}
    \beta(M_{\rm PBH}) = 
    \frac{\gamma}{\sqrt{2\pi\sigma_\delta^2}}
    \int_{\delta_{\mathrm{th}}}^{\infty}
    \exp\left(
    -\frac{\delta^2}{2\sigma_{\delta}^2}
    \right) 
    {\rm d}\delta
    =
    \gamma~\mathrm{erfc}\left(
    \frac{\delta_{\mathrm{th}}}{\sqrt{2}\sigma_{\delta}}
    \right)
    \,,
\end{align}
where 
\begin{align}
\sigma_\delta^2 (M_H)=\langle \delta^2 (t, {\bf x}) \rangle =\int_0^\infty W^2(k; R) \mathcal{P}_\delta(t, k)~\mathrm{d}(\ln k),
     \end{align}
is the variance of density contrast smoothed over $R$ using the window function $W(k;R)$.  In the numerical calculation, we adopt the Gaussian window function, $W^2(k; R)=\exp\left(-k^2R^2/2\right)$. Taking into account the thermal history and evolution of the universe after PBH formation, the PBH fraction of CDM today is given as \cite{Ando:2018qdb}
\begin{align}
    f_{\rm PBH}(M_{\rm PBH}) \equiv
    \frac{\Omega_{\rm PBH}}{\Omega_{\rm DM}} \simeq
    \left(
    \frac{\beta(M)}{1.6\times10^{-9}}
    \right)
    \left(
    \frac{10.75}{g_*(T)}
    \right)^{1/4}
    \left(
    \frac{0.12}{\Omega_{\rm DM}h^2}
    \right)
    \left(
    \frac{M_{\odot}}{M_{\rm PBH}}
    \right)^{1/2}
    \,.
\end{align}
The energy density of the stochastic GW background induced by the enhanced small-scale scalar perturbations \cite{Matarrese:1997ay,Mollerach:2003nq,Ananda:2006af,Baumann:2007zm} today is given by~\cite{Espinosa:2018eve, Kohri:2018awv, Domenech:2025bvr}
\begin{align}
    \Omega_{\mathrm{GW}}\left(\eta_0, k\right)=c_g \frac{\Omega_{r, 0}}{6} \int_0^{\infty} {\rm d} v \int_{|1-v|}^{1+v} {\rm d} u\left(\frac{4 v^2-\left(1+v^2-u^2\right)^2}{4 u v}\right)^2 \overline{\mathcal{I}^2(v, u)} \mathcal{P}_{\mathcal{R}}(k v) \mathcal{P}_{\mathcal{R}}(k u)\,,
\end{align}
with $ \Omega_{r,0} \simeq 5.38\times 10^{-5}$ the current radiation energy fraction at conformal time $\eta_{0}$ and $c_{g}\simeq 0.4$. In radiation domination era, the expression for the kernel $\overline{\mathcal{I}^2(v, u)}$ is given by
\begin{eqnarray}
    \begin{aligned}
\overline{\mathcal{I}^2(v, u)}=\frac{1}{2}\left[\frac{3\left(u^2+v^2-3\right)}{4 u^3 v^3}\right]^2 & {\left[\left(-4 u v+\left(u^2+v^2-3\right) \ln \left|\frac{3-(u+v)^2}{3-(u-v)^2}\right|\right)^2\right.} \\
& \left.+\pi^2\left(u^2+v^2-3\right)^2 \Theta(v+u-\sqrt{3})\right]\,.
    \end{aligned}
\end{eqnarray}

\subsection{Case $\gamma_m = \gamma_{4f} = 1$}

\begin{figure}[t!]
    \centering
    \includegraphics[width=0.58\linewidth]{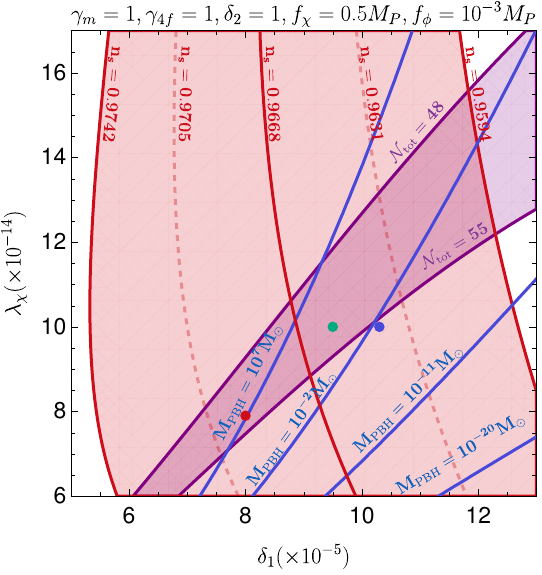}
    \caption{Parameter space in the $(\delta_1,\lambda_{\chi})$ plane where we have fixed $\gamma_m=1$, $\gamma_{4f}=1,$ $\delta_2=1$, $f_{\chi}=0.5 M_P$ and $f_{\phi}=10^{-3} M_P$. The red region corresponds to the constraint on the spectral index $n_s$. The central red line corresponds to the central value $n_s=0.9668$, while the dotted lines represent the $1\sigma$ contour and the external red lines the $2\sigma$ contour. The purple region represent the constraint on the number of $e$-folds $N_{\rm tot}=N_1+N_3$. The blue lines correspond to different masses for the PBH produced at the end of the first stage of inflation, which depend on $N_1$.}
    \label{fig:parameterspacev2}
\end{figure}

In order to have quantitative results for the PBH and GW production, we look for a parameter space satisfying the constraints on inflation \cite{Planck:2018jri}. We first consider the case where $\gamma_m=\gamma_{4f}=1$, values of the anomalous dimensions that are more natural in a composite model. To find a suitable parameter space, we require:
\begin{equation}
    n_s = 0.9668 \pm 0.0074,~~~~~~~~~~N_{\rm tot} \in [48,55].
    \label{inflationconstrains}
\end{equation}
 We fix some of the potential parameters according to \cite{Cacciapaglia:2023kat}:
 \begin{equation}
     f_{\chi}=0.5 M_P,~~~~~~~~~~~f_{\phi}=10^{-3} M_P,~~~~~~~~~~\delta_2=1.
 \end{equation}
  As the term depending on $\delta_3$ doesn't impact the inflationary dynamics of Stages $1$ and $3$, its value doesn't impact the parameter space, and we choose it as a tuning parameter around $\delta_3 \sim 10^{-12}$. By fixing those parameters, we now plot the parameter space constrained by Eq.~\eqref{inflationconstrains} in the $(\delta_1,\lambda_{\chi})$ plane, represented in Figure \ref{fig:parameterspacev2}. The red region corresponds to the parameter space satisfying the constraint on the spectral index $n_s$, with the red lines and dotted lines corresponding to the $1\sigma$ and $2\sigma$ contours. The purple region satisfies the constraint on the number of e-folds. The blue lines correspond to the mass of the PBH produced at the end of the first stage of inflation. From this parameter space, we choose a few benchmark parameters satisfying the inflationary constraints and corresponding to different PBH masses. They are represented in Figure \ref{fig:parameterspacev2} with blue, red, and green dots, with the values of the potential parameters for each benchmark parameter listed in Table \ref{tab:Benchmark}.  For a discussion of the impact of the recent Atacama Cosmology Telescope~\cite{AtacamaCosmologyTelescope:2025blo,AtacamaCosmologyTelescope:2025nti} measurements, we refer to Appendix~\ref{app:ACT}.

\begin{table}
    \centering
    \begin{tabular}{||c|cccccc||}
          \hline
          & $\lambda_{\chi}$ & $\delta_{1} $ & $\delta_{2}$ & $\delta_{3} $& $f_\chi (M_{P})$ & $f_{\phi} (M_P)$ \\
          \hline
          \hline 
         Set 1 (Blue)& $1 \times 10^{-13}$  & $1.03 \times 10^{-4}$& $1$&  $2.415\times10^{-12}$& $0.5 $ & $10^{-3}$\\
         Set 2 (Red)& $7.9 \times 10^{-14}$ & $8 \times 10^{-5}$ & $1$ & $4 \times 10^{-13}$& $0.5$ & $10^{-3}$\\
         Set 3 (Green)& $1 \times 10^{-13}$  & $9.5 \times 10^{-5}$& $1$& $2.3\times10^{-12}$& $0.5 $ &$10^{-3}$\\
         \hline
    \end{tabular}
    \caption{Benchmark parameters for $\gamma_{m} =\gamma_{4f} = 1$. Those parameters are chosen such that they satisfy both inflationary constraints. They correspond to different PBH masses produced at the end of stage $1$ of inflation.}
    \label{tab:Benchmark}
\end{table}
  
\begin{figure}[h]
    \centering
    \includegraphics[width=0.6\linewidth]{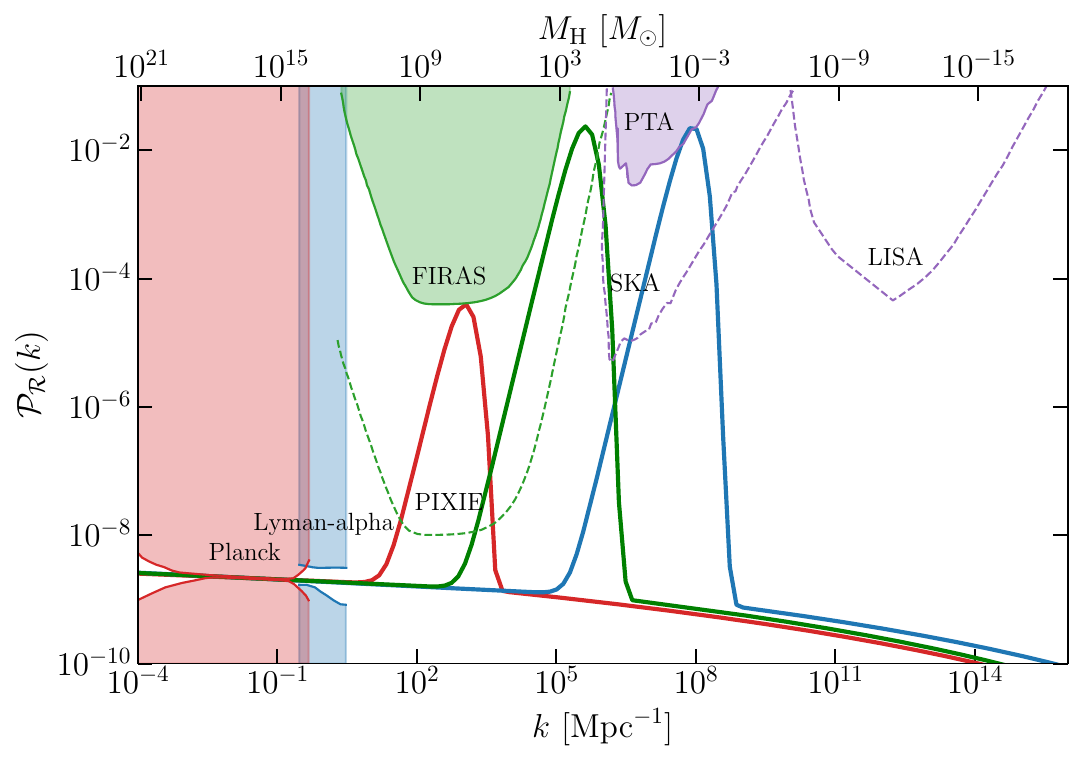}
    \caption{The $\mathcal{P}_{\mathcal{R}}(k)$ for benchmark parameters in Table \ref{tab:Benchmark}.  Various constraints on the curvature power spectrum~\cite{kavanagh2019} are also depicted, including Planck 2018~\cite{Planck:2018jri}, Lyman-$\alpha$~\cite{Bird:2010mp}, CMB $\mu$ distortions from the Far Infrared Absolute Spectrophotometer (FIRAS) experiment~\cite{Bianchini:2022dqh}, Pulsar Timing Arrays (PTAs)~\cite{Iovino:2024tyg,NANOGrav:2023hvm,EPTA:2023xxk, Reardon:2023zen,Xu:2023wog}, as well projected sensitivities of Primordial Inflation Explorer (PIXIE)~\cite{Chluba:2013pya, Abitbol:2017vwa}, Square Kilometre Array (SKA)~\cite{Weltman:2018zrl} and Laser Interferometer Space Antenna (LISA)~\cite{Baker:2019nia}.}
    \label{fig:powerspectrumparameterset}
\end{figure}

  The power spectrum for these parameters evaluated at the end of inflation $\mathcal{P}_{\mathcal{R}}(k,\mathcal{N}_{\mathrm{end}})$ is shown in Figure \ref{fig:powerspectrumparameterset}. All the curves exhibit a $k^{3}$ growth, with our benchmark points sitting from PTA scales to CMB distortions. This sharp peak in  $\mathcal{P}_{\mathcal{R}}(k,\mathcal{N}_{\mathrm{end}})$ can grow to values that correspond to copious PBH production and a potentially observable SIGW background.

\begin{figure}[h]
    \centering
    \includegraphics[width=0.6\linewidth]{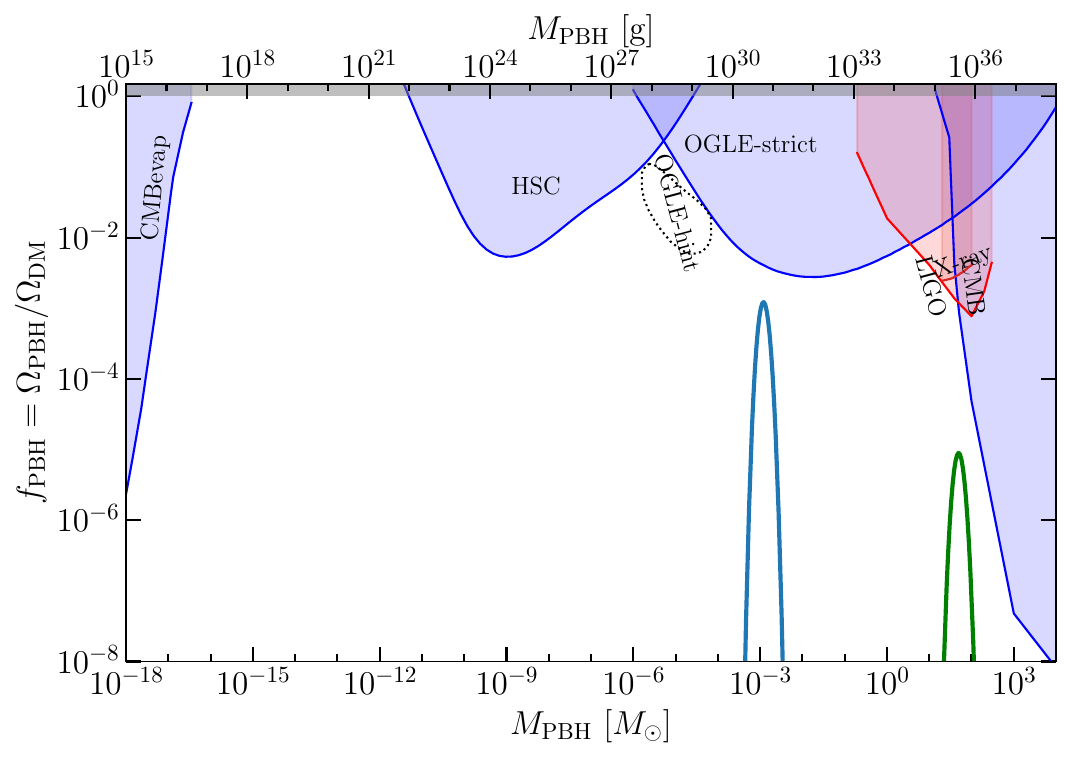}
    
    \caption{The PBH fraction $f_{\mathrm{PBH}} = \Omega_{\mathrm{PBH}} / \Omega_{\mathrm{DM}}$ for benchmark parameters in Table \ref{tab:Benchmark}, with constraints from \cite{kavanagh2019} overlaid. Note that Set 3 is located beyond the range of this plot.}
    \label{fig:fPBH}
\end{figure}

\begin{figure}[h]
    \centering
    \includegraphics[width=0.6\linewidth]{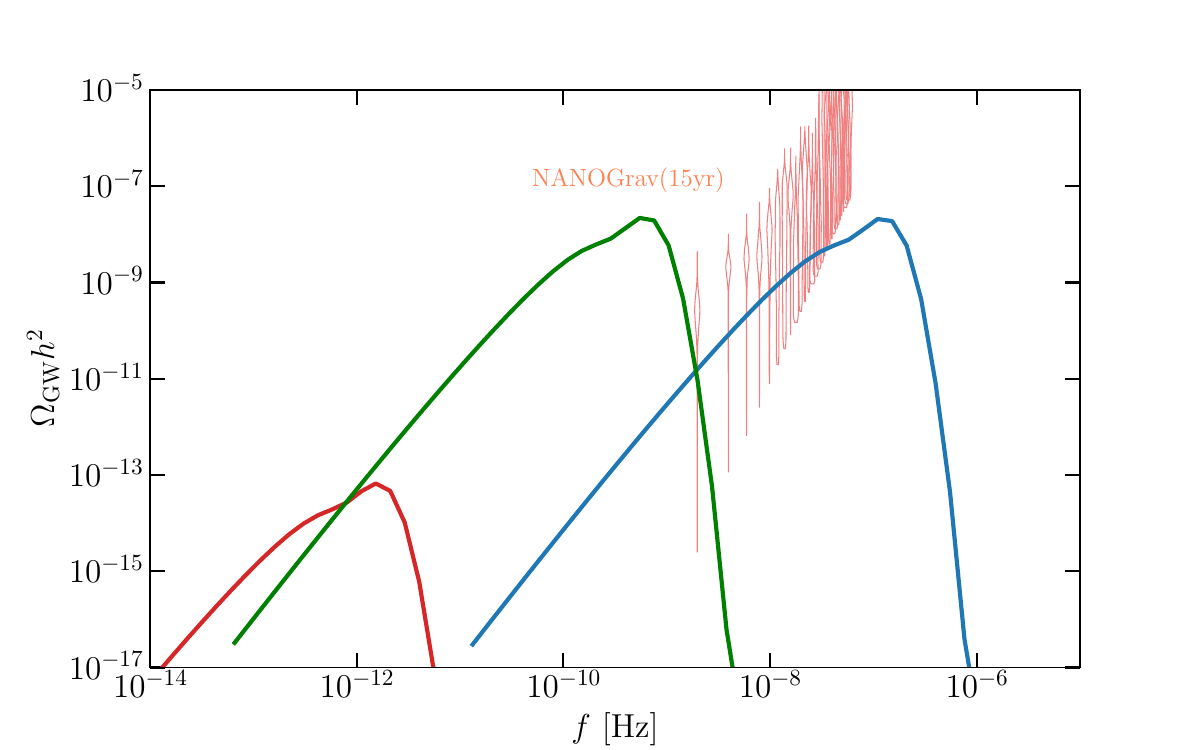}
    \caption{The stochastic GW energy spectrum $\Omega_{\mathrm{GW}}$ for benchmark parameters in Table \ref{tab:Benchmark}. The violin (light-red) curves correspond to the reported gravitational wave background in the NANOGrav 15yr dataset~\cite{NANOGrav:2023hvm}. } 
    \label{fig:SIGW}
\end{figure}

  The PBH abundance and the stochastic GW background, evaluated for the same benchmark parameters, are shown in Figure \ref{fig:fPBH} and Figure \ref{fig:SIGW}, respectively. As depicted in Figure \ref{fig:parameterspacev2}, the PBH masses lie in a region incompatible with explaining the entirety of dark matter, which is disfavoured by microlensing and dynamical constraints. The GW background, however, can cover frequency domains probed by PTAs, with the possibility of the enhanced GW spectrum being compatible with the reported PTA background~\cite{NANOGrav:2023gor,NANOGrav:2023hde,EPTA:2023fyk, Tagliazucchi:2023dai}.


\subsection{Case $\gamma_m > 1$ and $\gamma_{4f} > 1$}
\label{sec:gammage1}

\begin{figure}[h]
    \centering
    \includegraphics[width=0.6\linewidth]{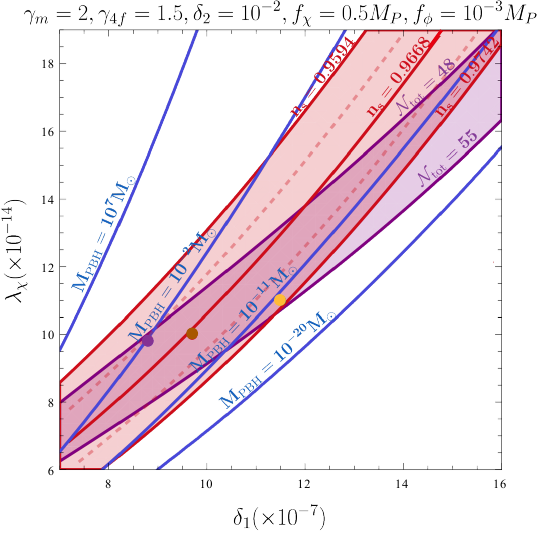}
    \caption{Parameter space in the $(\delta_1,\lambda_{\chi})$ plane for $\delta_2=10^{-2}$, $f_{\chi}=0.5~M_P$, $f_{\phi}=10^{-3}~M_P$, $\gamma_m=2$ and $\gamma_{4f}=1.5.$ As in Figure \ref{fig:parameterspacev2}, the red region represents the constraint on the spectral index $n_s = 0.9668 \pm 0.0074$ up to $2\sigma$ while the purple region constrains the number of e-folds $N_{\rm tot} \in [48,55] $. The blue lines correspond to the PBH mass, which depends on $N_1$. Compared to the case where $\gamma_m=\gamma_{4f}=1$, we notice that the constraints on inflation allow PBH masses around asteroid masses, $M_{\rm PBH} \sim 10^{-11} M_{\odot}$.}
    \label{fig:Parameterspacegamma2_1.5}
\end{figure}

\begin{figure}[h]
    \centering
    \includegraphics[width=0.6\linewidth]{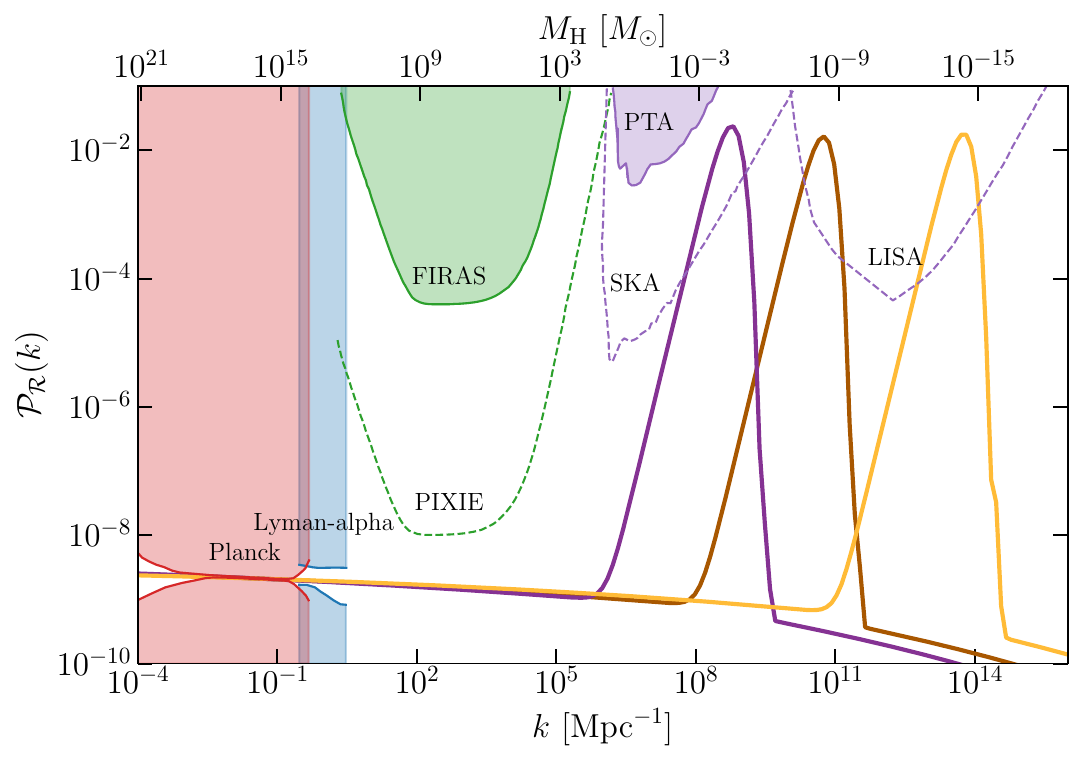}
    \caption{The $\mathcal{P}_{\mathcal{R}}(k)$ for benchmark parameters in Table \ref{tab:Benchmark2}. Constraints on $\mathcal{P}_{\mathcal{R}}(k) $ are identical to Figure~\ref{fig:powerspectrumparameterset}.}
    \label{fig:powerspectrumparameterset2}
\end{figure}

\begin{figure}[h]
    \centering
    \includegraphics[width=0.6\linewidth]{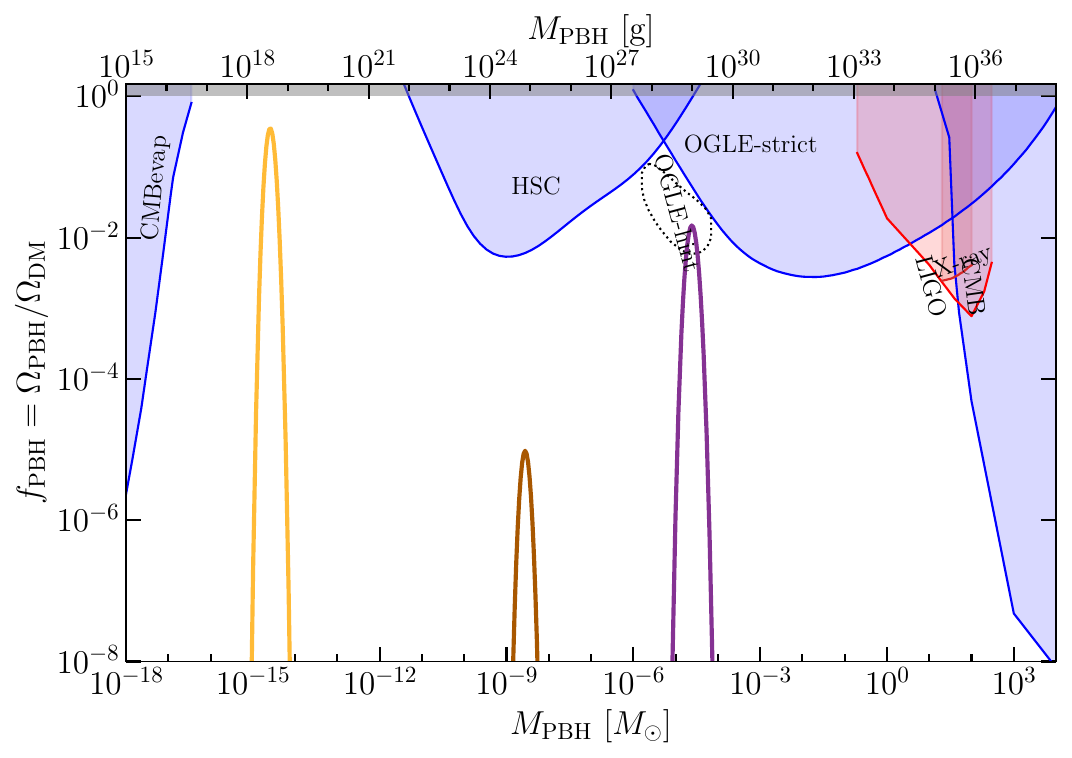}
    \caption{The PBH fraction $f_{\mathrm{PBH}} = \Omega_{\mathrm{PBH}} / \Omega_{\mathrm{DM}}$ for benchmark parameters in Table \ref{tab:Benchmark2}, with constraints identical to Figure~\ref{fig:fPBH}.}
    \label{fig:fPBH2}
\end{figure}

\begin{figure}[h]
    \centering
    \includegraphics[width=0.6\linewidth]{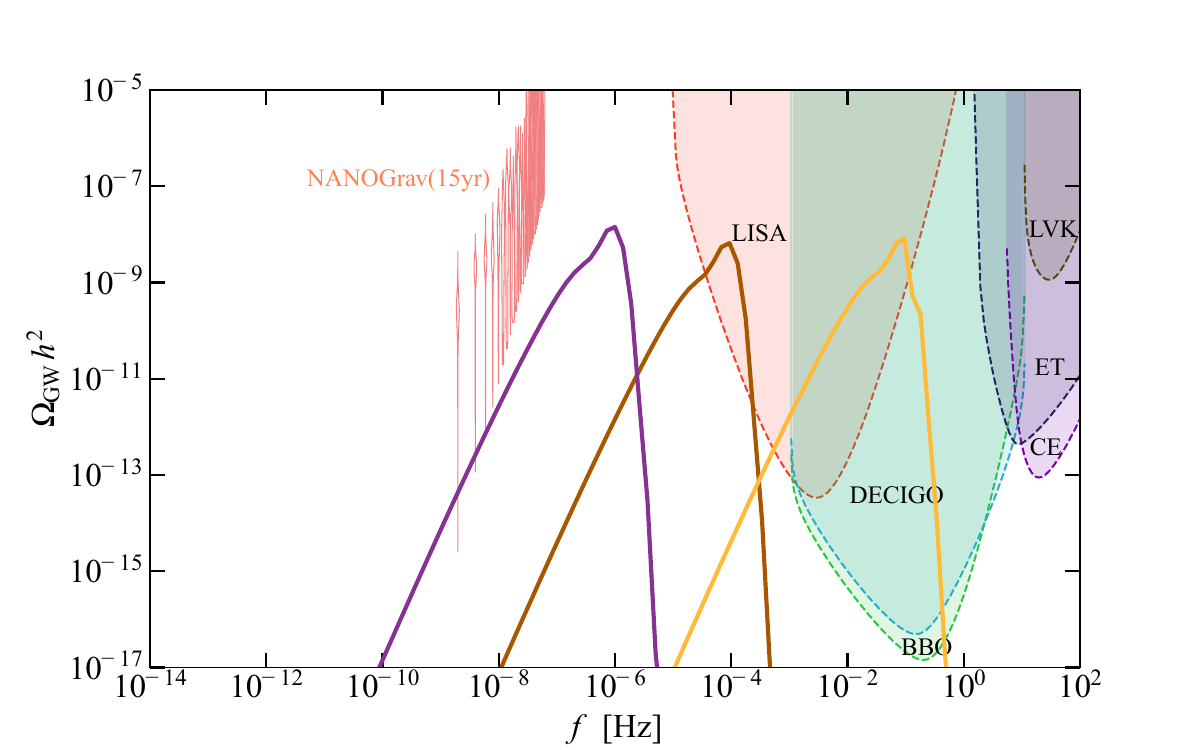}
    \caption{The stochastic GW energy spectrum $\Omega_{\mathrm{GW}}$ for benchmark parameters in Table \ref{tab:Benchmark2} with PTA results and sensitivities for current LIGO-VIRGO-KAGRA (LVK) detector network~\cite{LIGOScientific:2014pky,VIRGO:2014yos,KAGRA:2018plz} and projected sensitivities for future GW observatories,  LISA~\cite{LISA:2017pwj}, Deci-hertz Interferometer Gravitational wave Observatory (DECIGO)~\cite{Kawamura:2011zz}, and Big Bang Observer (BBO)~\cite{Crowder:2005nr} Einstein Telescope (ET)~\cite{Punturo:2010zz}, and Cosmic Explorer (CE)~\cite{Reitze:2019iox}.} 
    \label{fig:SIGW2}
\end{figure}

We now alleviate the restriction on the anomalous dimensions, and consider scenarios where $\gamma_{m}$ and $\gamma_{4f}$ take values larger than unity. The parameter space satisfying the constraints on inflation can be found in a similar way to that in the previous section. As a concrete example, Figure~\ref{fig:Parameterspacegamma2_1.5} depicts the parameter space for $\gamma_{m} = 2$ and $\gamma_{4f} = 1.5$, with the corresponding benchmark parameters for this scenario in Table~\ref{tab:Benchmark2}. The power spectrum for these parameters is shown in Figure \ref{fig:powerspectrumparameterset2}, and the PBH abundance and the stochastic GW background for these parameters are shown in Figure \ref{fig:fPBH2} and Figure \ref{fig:SIGW2}, respectively.

In this region, the correlation between $n_{s}$, $N_{\mathrm{tot}}$ and $M_{\mathrm{PBH}}$ differ from the $\gamma_{m} = \gamma_{4f} =1 $ case, as lighter $M_{\mathrm{PBH}}$ (correspondingly a longer $N_{\mathrm{tot}}$) gives a larger $n_{s}$ value. The linear dilaton potential term defined by $\gamma_{m} = 2$ provides greater compatibility with CMB observations~\cite{Cacciapaglia:2023kat}, therefore the region satisfying CMB observables and $N_{\mathrm{tot}} = 48 \sim 55$ allows much lighter PBHs to be produced, opening the possibility for these PBHs to constitute the whole dark matter density. One numerical example is provided by Set 3 of our benchmark parameters. The compatibility with lighter PBHs may also be the explanation for the OGLE microlensing events~\cite{Niikura:2019kqi}. The SIGWs span across higher frequencies where future GW observatories, including LISA~\cite{LISACosmologyWorkingGroup:2024hsc, LISACosmologyWorkingGroup:2025vdz}, {provide strong prospects to test this scenario.}

\begin{table}
    \centering
    \begin{tabular}{||c|cccccc||}
          \hline
          & $\lambda_{\chi}$ & $\delta_{1} $ & $\delta_{2}$ & $\delta_{3} $& $f_\chi (M_{P})$ & $f_{\phi} (M_P)$ \\
          \hline
          \hline 
         Set 1 (Purple)& $9.8 \times 10^{-14}$  & $8.8 \times 10^{-7}$& $10^{-2}$&  $2.118\times10^{-14}$& $0.5 $ & $10^{-3}$\\
         Set 2 (Brown)& $1\times 10^{-13}$ & $9.7 \times 10^{-7}$ & $10^{-2}$ & $2.683 \times 10^{-14}$& $0.5$ & $10^{-3}$\\
         Set 3 (Yellow)& $1.1 \times 10^{-13}$  & $1.15 \times 10^{-6}$& $10^{-2}$& $2.878\times10^{-14}$& $0.5 $ &$10^{-3}$\\
         \hline
    \end{tabular}
    \caption{Benchmark parameters for $\gamma_{m} =2,~\gamma_{4f} = 1.5$. Those parameters are chosen such that they satisfy both inflationary constraints. They correspond to different PBH masses produced at the end of stage $1$ of inflation.}
    \label{tab:Benchmark2}
\end{table}


\section{Conclusions}
In this work, we explored the production of primordial black holes and stochastic gravitational waves within an inflationary scenario originating from a composite sector. By considering a confining gauge theory that couples to fundamental fermions, we naturally induce the shift symmetry needed in the inflaton potential through the dilaton, along with additional pions. We include a $\mathbb{Z}_{2}$ breaking term in the pion potential, in order to prevent vacuum degeneracy and ensure domain wall decay. As a consequence, the inflation trajectory features a non-trivial period of tachyonic isocurvature perturbation enhancement during slow-roll inflation. Within consistency with composite theories, we show that this tachyonic instability can last for $\mathcal{O}(5) $ e-folds, exponentially enhancing the isocurvature perturbations, which then source a growth in the curvature perturbation. 

The numerical predictions crucially depend on the anomalous dimensions of the pion operators, which control the coupling of the dilaton to the pions. For anomalous dimensions $\gamma_{m} = \gamma_{4f} = 1$, which are naturally expected from walking composite dynamics, we demonstrate that there is a tight correlation between the total number of e-folds and the e-folding time of the start of the tachyonic instability, hence restricting the mass of PBHs to be heavier than $10^{-3} M_{\odot} $ within $2\sigma$ consistency with CMB observables. We also show that these parameters induce a stochastic gravitational wave background compatible with the frequency range of recent PTA observations. We also show that relaxing the conditions on the anomalous dimensions, hence allowing values above unity, the mass of the PBHs can be significantly reduced, allowing them to constitute the whole dark matter density. Furthermore, the stochastic gravitational wave background appears at higher frequencies, within reach of future observatories like LISA and DECIGO. Overall, we show that the underlying structure of the composite theory is directly related to the properties of produced primordial black holes and gravitational waves. 

There are several interesting points that can be further elaborated within this setup. Within the inflationary potential, we introduced the $\mathbb{Z}_{2}$ breaking term in order to prevent domain wall stability. However, the actual dynamics of the domain walls during their decay might also lead to interesting phenomenology, such as dark matter production, and a gravitational wave background that differs in frequency and spectral behaviors. We leave the exploration of these aspects for future work.

\acknowledgments
We thank Misao Sasaki and Juhoon Son for useful discussions. 
This work was supported by the National Research Foundation of Korea(NRF) grant funded by the Korea government (MSIT) (RS-2023-00283129), (RS-2024-00340153) and Yonsei internal grant for Mega-science (2023-22-048). [SCP, DYC].
This work was also supported in part by JSPS KAKENHI Grant Nos. JP20H05853 and JP24K00624, and Shanghai Pujiang Program 24PJA134 [XW, YZ], by Forefront Physics and Mathematics Program to Drive Transformation (FoPM), a World-leading Innovative Graduate Study (WINGS) Program, the University of Tokyo [XW], by the Fundamental Research Funds for the Central Universities, and the Project 12475060 and 12047503 supported by NSFC, and Project 24ZR1472400 sponsored by Natural Science Foundation of Shanghai [YZ].
Kavli IPMU is supported by World Premier International Research Center Initiative (WPI), MEXT, Japan.
We also acknowledge partial support from the France-Korea PHC STAR 2024 project ``The top quark and the Higgs boson as windows on beyond the standard model physics''.


\appendix

\section{Domain wall - formation and decay : (B)SM radiation and gravitational waves} \label{app:domainwalls}

The $\mathbb{Z}_{2} $ breaking term may also be responsible for the decay of domain walls, which could also give GW signatures as well. We follow the procedure from Ref~\cite{Ferreira:2022zzo} and apply it to our model. Assuming that domain walls decay at some specific time $t_{*}$, the energy density fraction of domain walls at that time (correspondingly temperature, assuming a radiation-dominated universe) will give:
\bea
\alpha_{*} \equiv \frac{\rho_{\text{DW}}}{\rho_{\text{cr}}} \simeq c \sqrt{\frac{g_{*}(T_{*})}{10.75}} \left(\frac{\sigma^{1/3}}{10^{5}~\text{GeV}} \right)^{3} \left( \frac{10~\text{MeV}}{T_{*}}\right)^{2},
\eea 
where $\rho_{\text{DW}} = c \sigma H\simeq \Delta V$ and $\sigma$ is the surface energy density of the domain wall. For a finite potential gap $\Delta V$, the annihilation temperature takes the form:
\begin{align}
    T_{\star} \simeq 5 \mathrm{MeV}\left(\frac{10.75}{g_*\left(T_{\star}\right)}\right)^{\frac{1}{4}}\left(\frac{\Delta V^{1 / 4}}{10 \mathrm{MeV}}\right)^2\left(\frac{10^5 \mathrm{GeV}}{(c\sigma)^{1 / 3}}\right)^{\frac{3}{2}}.
\end{align}
At the end of inflation, the potential gap induced by $\delta_3$ at the end of inflation reads
\begin{align}
    \left|{\Delta V}\right|&=\lambda_{\chi}\delta_3f_\chi^4\left(\frac{\chi_0}{f_{\chi}}\right)^{3-\gamma_{m}}\sin\frac{\phi_0}{f_\phi}\\
    &\approx\left(5.0 \times 10^{15} \rm MeV\right)^4\left(\frac{\lambda_{\chi}}{10^{-13}}\right)\left(\frac{\delta_3}{2.3\times10^{-12}}\right)\left(\frac{f_\chi}{0.5 M_{P}}\right)^4,
\end{align}
where we used $M_{P}=1.22\times 10^{22}\rm MeV$ and we chose the values $\delta_1 = 10^{-4}$ and $\delta_2=1$, see Section \ref{sec:gamma1}.
The domain wall surface energy density at the end of inflation can be given by
\begin{align}
   c\sigma \simeq \frac{\Delta V}{H}&\approx\frac{M_{\rm pl}\Delta V}{\sqrt{V_0/2}}={4\sqrt{2}}\sqrt{\lambda_{\chi}}\delta_3 f_{\chi}^2\\&\sim(1.2\times 10^{16} \mathrm{MeV})^3\left(\frac{\lambda_{\chi}}{10^{-13}}\right)^{1/2}\left(\frac{\delta_3}{2.3\times10^{-12}}\right)\left(\frac{f_\chi}{0.5 M_{P}}\right)^2,
\end{align}
if we take $\epsilon_{H}=1$ at the end of inflation.
Thus the corresponding decay temperature $T_*$ and $\alpha_*$ are given by
\begin{align}
    &T_{\star} \simeq 9.5\times 10^{17} \mathrm{MeV}\left(\frac{10.75}{g_*\left(T_{\star}\right)}\right)^{\frac{1}{4}}\left(\frac{\lambda_{\chi}}{10^{-13}}\right)^{1/4}\left(\frac{f_\chi}{0.5 M_{P}}\right),\\
      &\alpha_{*}\simeq 1.9\times 10^{-10}\sqrt{\frac{g_*(T_*)}{10.75}}\left(\frac{\delta_3}{2.3\times10^{-12}}\right).
\end{align}
Since most of the GWs emitted during DW evaporation at the frequency featured by comoving scale of the domain wall, we estimate the peak of the GW spectrum today by
\begin{align}
    f_{p}^0 \simeq 9.5\times10^{7}\mathrm{Hz}\left(\frac{g_*\left(T_{\star}\right)}{10.75}\right)^{-\frac{1}{12}}\left(\frac{\lambda_{\chi}}{10^{-13}}\right)^{1/4}\left(\frac{f_\chi}{0.5 M_{P}}\right)\,,
\end{align}
which is too high for existing GW observatories. However, these high frequency gravitational waves can potentially be accessible with ultra-high-frequency gravitational wave detectors~\cite{Aggarwal:2025noe}.

\section{Parameter space compatibility with ACT observations} \label{app:ACT}

In this appendix we address the CMB observables from the 6th data release (DR6) of the Atacama Cosmology Telescope~\cite{AtacamaCosmologyTelescope:2025blo,AtacamaCosmologyTelescope:2025nti}. A joint Planck 2018 + ACT DR6 + DESI Y1 analysis reports a spectral index $n_{s}(k^*=0.05~ {\rm Mpc}^{-1}) = 0.9743 \pm 0.0034$, values larger than previous Planck 2018~\cite{Planck:2018jri} reports, with the mean values of the two reports incompatible at the $2\sigma$ level. This larger $n_s$ value can alter compatibility among inflationary models, which for our consideration may change the overall parameter region compatible with PBH dark matter and observable GWs. 

\begin{figure}[t!]
    \centering
    \includegraphics[width=0.365\linewidth]{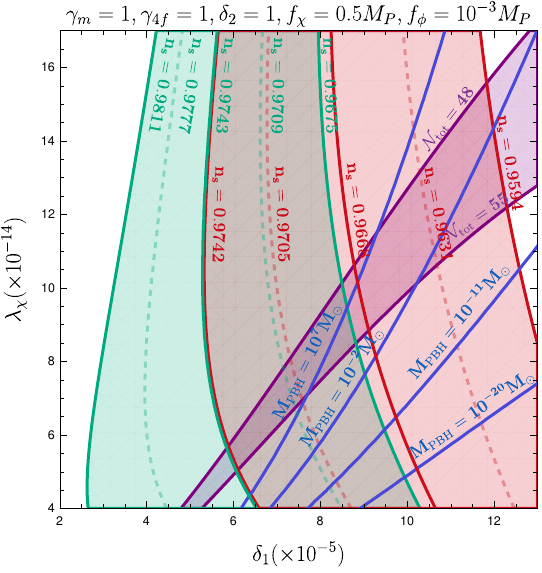}
    \includegraphics[width=0.55\linewidth]{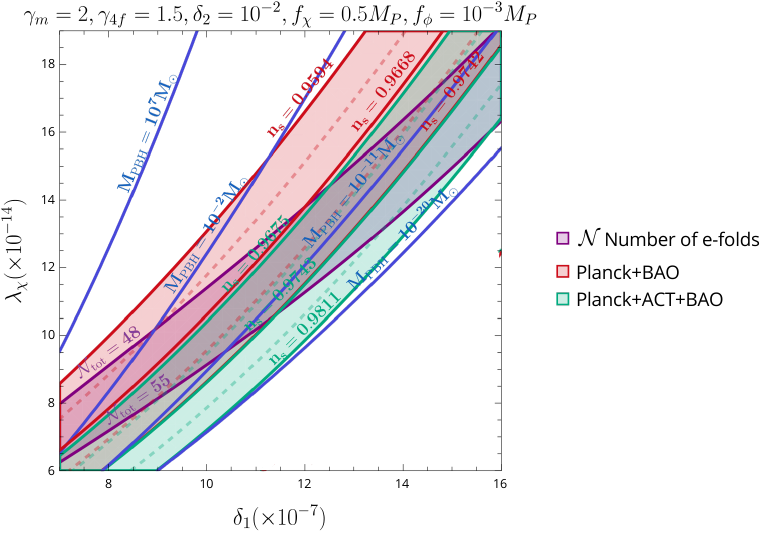}
    \caption{Parameter space in the $(\delta_1,\lambda_{\chi})$ plane for $\gamma_m=1$, $\gamma_{4f}=1,$ $\delta_2=1$, $f_{\chi}=0.5 M_P$ and $f_{\phi}=10^{-3} M_P$ (top), and $\gamma_m=2$, $\gamma_{4f}=1.5$ $\delta_2=1$, $f_{\chi}=0.5 M_P$ and $f_{\phi}=10^{-3} M_P$ (bottom). The color scheme follows that of Figure~\ref{fig:parameterspacev2} and Figure~\ref{fig:Parameterspacegamma2_1.5}, with the additional light green region corresponding to the ACT results~\cite{,AtacamaCosmologyTelescope:2025nti}.  }
    \label{fig:parameterspaceACT}
\end{figure}
In Figure~\ref{fig:parameterspaceACT} we show the parameter space in the $(\delta_1,\lambda_{\chi})$ plane for both Planck 2018 results and ACT results. For $\gamma_{m} = \gamma_{4f}= 1$, due to the shifted $n_{s}$ value the PBH mass that is compatible with CMB observables within an inflation duration of 55 efolds shifts to higher values, with the compatible SIGW background frequencies also lower than PTA observatories. In contrast, for $\gamma_{m}=2$, $\gamma_{4f} = 1.5$, the different tendency in the $n_s$ parameter actually allows within $1\sigma$ compatibility with the CMB observations, allowing PBH production within the entire PBH dark-matter window. 

\bibliographystyle{JHEP}
\bibliography{ref}
\end{document}